\documentclass[%
 aip,
 amsmath,amssymb,
 reprint,groupedaddress
]{revtex4-1}

\DeclareUnicodeCharacter{2212}{-}
\usepackage{graphicx}
\usepackage{dcolumn}
\usepackage{algorithm}
\usepackage{algpseudocode}
\usepackage{amsmath}
\usepackage{float}
\usepackage{comment}
\usepackage{threeparttable}

\usepackage[utf8]{inputenc}
\usepackage[T1]{fontenc}
\usepackage{etoolbox}
\usepackage{textcomp}

\usepackage{fancyhdr}

\usepackage{xcolor}

\newcommand{\nick}[1]{\textcolor{black}{#1}}

\makeatletter
\def\@email#1#2{%
 \endgroup
 \patchcmd{\titleblock@produce}
  {\frontmatter@RRAPformat}
  {\frontmatter@RRAPformat{\produce@RRAP{*#1\href{mailto:#2}{#2}}}\frontmatter@RRAPformat}
  {}{}
}%
\makeatother
\begin{document}

\preprint{AIP/123-QED}

\title[]{Real-time nuclear--electronic orbital time-dependent density functional theory with a constrained traveling proton basis}
\author{Nicholas J. Boyer}

\author{Sharon Hammes-Schiffer*}%
 \email{shs566@princeton.edu}
\affiliation{ 
Department of Chemistry, Princeton University, Princeton, New Jersey 08544, USA
}%

\date{\today}

\begin{abstract}
Nuclear quantum effects and non-Born--Oppenheimer effects play a vital role in many chemical and biological processes, motivating the incorporation of such effects into dynamical simulations. In real-time nuclear--electronic orbital time-dependent density functional theory (RT-NEO-TDDFT), the electronic and nuclear densities are propagated numerically in time according to the time-dependent Schrödinger equation. In this framework, specified protons are treated quantum mechanically on the same level as the electrons. The classical nuclei can be propagated on the instantaneous NEO vibronic surface using Ehrenfest dynamics. A traveling proton basis (TPB) can be used to describe the dynamics of moving protons in conjunction with Gaussian-type protonic and electronic basis sets for each quantum proton. Herein, we present a constrained TPB (c-TPB) approach that ensures each protonic basis function center coincides with the corresponding proton position expectation value during the dynamics. This approach produces accurate nuclear--electronic quantum dynamics and rigorously conserves energy.  We demonstrate the accuracy and stability of this approach for computing molecular vibrational frequencies as well as simulating excited-state intramolecular proton transfer and double proton transfer in the \textit {o}-hydroxybenzaldehyde and [2,2$'$-bipyridyl]-3,3$'$-diol molecules. These applications show that the c-TPB method provides accurate dynamics, conserves energy, and is computationally efficient.
\end{abstract}

\maketitle

\section{Introduction}

Nuclear quantum effects, such as zero-point energy, nuclear delocalization, and nuclear tunneling, are important throughout chemistry and biology.\cite{10.1063/1.2827006, doi:10.1021/acs.chemrev.5b00674, doi:10.1021/acs.jpclett.1c01987,doi:10.1126/sciadv.adw4798} Examples of processes that are strongly influenced by nuclear quantum effects, as well as nonadiabatic or non-Born--Oppenheimer effects, include proton transfer, proton-coupled electron transfer, and proton-coupled energy transfer.\cite{klinmanHydrogenTunnelingLinks2013, layfield_hydrogen_2014,VARDIKILSHTAIN201518,doi:10.1021/acs.accounts.5c00119, hammes-schifferProtonCoupledElectronEnergy2026}
In addition, experimental observables such as kinetic isotope effects, vibrational spectra, and NMR spectra are affected by nuclear quantum effects. A wide range of computational methods have been developed to incorporate nuclear quantum effects and nonadiabatic effects in simulations of chemical and biological systems.\cite{mantheWavepacketDynamicsMulticonfiguration1992,martinezMultiElectronicStateMolecularDynamics1996,jangPathIntegralCentroid1999,abediExactFactorizationTimeDependent2010,doi:10.1021/ct100716c,habershonRingPolymerMolecularDynamics2013a, richingsQuantumDynamicsSimulations2015,curchodInitioNonadiabaticQuantum2018,10.1063/1.5024869,liRealTimeTimeDependentElectronic2020, doi:10.1021/acs.jpclett.3c01848} 

One strategy that balances accuracy and computational efficiency is the nuclear electronic orbital (NEO) approach, which treats nuclei quantum mechanically at the same level as the electrons.\cite{10.1063/1.1494980, doi:10.1021/acs.chemrev.9b00798, hammes-schiffer_nuclearelectronic_2021}
This approach removes the Born--Oppenheimer separation between the electrons and quantum nuclei. A variety of NEO methods have been developed, including NEO coupled-cluster,\cite{pavosevic_multicomponent_2019,fowler_t_2022, 10.1063/5.0303185} perturbation theory,\cite{swalina_impact_2005,pavosevic_multicomponent_2020, hasecke_local_2024} multiconfigurational,\cite{10.1063/1.1494980, fajen_multicomponent_2021,doi:10.1021/acs.jpclett.4c00805} and multireference wavefunction methods,\cite{malbon_nuclearelectronic_2025,doi:10.1021/acs.jpclett.5c01657} as well as NEO density functional theory \cite{doi:10.1021/jp0704463,yang_development_2017,brorsen_multicomponent_2017} and multistate density functional theory.\cite{yuNuclearElectronicOrbitalMultistate2020,doi:10.1021/acs.jctc.2c00938,doi:10.1021/acs.jctc.4c00737}
 
Real-time NEO methods allow the simulation of vibrational spectra and chemical dynamics.\cite{doi:10.1021/acs.jpclett.0c00701, doi:10.1021/acs.accounts.1c00516,doi:10.1021/acs.jpclett.4c00805} In particular, real-time NEO time-dependent density functional theory (RT-NEO-TDDFT) uses the NEO Kohn-Sham Hamiltonian to propagate the electronic and nuclear densities in time.\cite{doi:10.1021/acs.jpclett.0c00701} Often only the protons are treated quantum mechanically. The other nuclei can be propagated classically using Ehrenfest dynamics, thereby including nonadiabatic effects between the classical and quantum subsystems.\cite{10.1063/5.0031019, zhao_excited_2021} These approaches have been used to explore photoinduced proton transfer in the gas phase,\cite{zhao_excited_2021} in solution,\cite{wildman_solvated_2022} at surfaces,\cite{xu_first-principles_2023} and in proteins,\cite{chow_nuclear_2024} as well as magnetic field effects\cite{doi:10.1021/acs.jctc.5c00273} and polaritonic chemistry.\cite{li_semiclassical_2022,doi:10.1021/acs.jctc.5c00911}

 A challenge of RT-NEO methods is the treatment of the Gaussian-type basis sets associated with the quantum nuclei, particularly when the protons move significantly during the dynamics. Previous RT-NEO-TDDFT studies have proposed several different approaches for treating the basis sets associated with the quantum protons.\cite{10.1063/5.0031019,10.1063/5.0230570,10.1063/5.0255984} One option is a fixed proton basis (FPB), where each quantum proton is represented by a set of fixed electronic and protonic basis sets centered at pre-specified spatial coordinates. However, this method can become very expensive, as it requires a large number of basis functions if the proton moves far from its initial position. Additionally, {\it a priori} knowledge of the proton pathway is required in order to determine the locations of the fixed protonic basis function centers before the dynamics. 
 
 An alternative approach is to utilize a traveling proton basis (TPB) that moves with the proton. Several variants of the TPB method have been explored in the literature.\cite{10.1063/5.0031019,zhao_excited_2021,10.1063/5.0255984,10.1063/5.0230570} The original TPB approach, in which the protonic basis function center moves according to classical equations of motion, has been found to provide sufficiently accurate dynamics for the  systems studied,\cite{10.1063/5.0031019, zhao_excited_2021} but it does not rigorously conserve energy. To address these limitations, an extended Lagrangian TPB method was developed.\cite{10.1063/5.0230570} This TPB approach conserves energy but may become unstable at longer times because it does not ensure that the protonic basis function center remains close to the expectation value of the proton position operator.

Herein, we present a solution to this problem using a constrained TPB (c-TPB) approach that both conserves energy and provides accurate dynamics in a stable and efficient manner. This approach is based on an extended Lagrangian that is used to derive the dynamical equations for the electronic and protonic densities, as well as the classical nuclei and protonic basis function centers. In addition, Lagrangian constraints are applied to ensure that the quantum proton position expectation value and corresponding protonic basis function center coincide during the dynamics for each quantum proton. The resulting RT-NEO-TDDFT c-TPB Ehrenfest dynamics approach allows the simulation of vibrational spectra and nonequilibrium nuclear--electronic quantum dynamics in a stable and computationally efficient manner.

An outline of this paper is as follows. In Section II, we derive the theoretical formulation and present the algorithm used to propagate the electronic and protonic densities, as well as the classical nuclei and protonic basis function centers. Section III provides the computational details for the applications to molecular systems.  Section IV presents the results of these applications, including vibrational frequency calculations as well as single and double proton transfer dynamics. Specifically, we compute the vibrational frequencies for molecules with a single quantum proton, HCN, HNC, and FHF$^-$, and with multiple quantum protons, H$_2$O, HCOOH, H$_2$CO, and H$_2$. We then simulate excited-state intramolecular proton transfer for single proton transfer in the \textit{o}-hydroxybenzaldehyde (oHBA) molecule and double proton transfer in the [2,2$'$-bipyridyl]-3,3$'$-diol (BP(OH)$_2$) molecule. These examples showcase the accuracy, stability, and efficiency of the RT-NEO-TDDFT c-TPB Ehrenfest dynamics method for simulating vibrational spectra and proton transfer reactions.

\section{Theory}

In this section, we review the original TPB method and present the c-TPB method. Here, we derive the TPB methods for the case of only one quantum proton. The derivations of these equations for the general case of multiple quantum protons are given in the Supplementary Material, and we provide these more general equations at the end of this section. For molecular systems, proton orbitals are sufficiently localized such that the quantum protons can be treated as distinguishable particles using the nuclear Hartree product representation. \cite{auer_localized_2010,10.1063/5.0308634} The proton--proton exchange terms have been shown to be eight to ten orders of magnitude smaller than their electronic counterparts. \cite{doi:10.1021/acs.chemrev.9b00798,10.1063/5.0308634} Thus, the extension of the TPB equations to multiple quantum protons is straightforward.

\subsection{RT-NEO-TDDFT with a Fixed Proton Basis}

The NEO Hamiltonian includes the kinetic energies of the electrons and quantum protons and the Coulomb interactions among the electrons, quantum protons, and classical nuclei. The RT-NEO-TDDFT approach expresses the reference nuclear--electronic wavefunction as the product of an electronic and a protonic Slater determinant, which are composed of electronic and protonic orbitals, respectively. As mentioned above, the protonic Slater determinant can be replaced by a protonic Hartree product. For a single quantum proton, the protonic part is simply a single proton orbital. Substituting this nuclear--electronic wavefunction into the time-dependent Schr\"odinger equation leads to two sets of coupled equations expressed in atomic units as
\begin{equation}
i \frac{\partial }{\partial t} \psi^{\mathrm{p}}(\mathbf{r^{\mathrm{p}}}, t) = \hat{F}^{\mathrm{p}}(\mathbf{r^{\mathrm{p}}}, t) \psi^{\mathrm{p}}(\mathbf{r^{\mathrm{p}}}, t)
\label{eq:TDSE}
\end{equation} 
\begin{equation}
i \frac{\partial }{\partial t} \psi^{\mathrm{e}}_n(\mathbf{r^{\mathrm{e}}}, t) = \hat{F}^{\mathrm{e}}(\mathbf{r^{\mathrm{e}}}, t) \psi^{\mathrm{e}}_n(\mathbf{r^{\mathrm{e}}}, t)
\label{eq:TDSE2}
\end{equation} 
\noindent where $\psi^{\mathrm{p}}(\mathbf{r}^{\mathrm{p}}, t)$ is the protonic orbital and $\psi_n^{\mathrm{e}}(\mathbf{r}^{\mathrm{e}}, t)$ are the electronic orbitals. Moreover, $\hat{F}^{\mathrm{p}}(\mathbf{r^{\mathrm{p}}}, t)$ is the one-proton Kohn-Sham operator, and $\hat{F}^{\mathrm{e}}(\mathbf{r^{\mathrm{e}}}, t)$ is the one-electron Kohn-Sham operator, defined as
\begin{equation}
\hat{F}^{\mathrm{p}}(\mathbf{r^{\mathrm{p}}}, t) = \frac{-1}{2 m_{\mathrm{p}}}
\nabla^2_{\mathbf{r^{\mathrm{p}}}} +U_{\mathrm{eff}}^{\mathrm{p}}(\mathbf{r^{\mathrm{p}}}, t)\end{equation}
\begin{equation}
\hat{F}^{\mathrm{e}}(\mathbf{r^{\mathrm{e}}}, t) = \frac{-1}{2 }
\nabla^2_{\mathbf{r^{\mathrm{e}}}} +U_{\mathrm{eff}}^{\mathrm{e}}(\mathbf{r^{\mathrm{e}}}, t) \hspace{0.5em}.\end{equation}

\noindent Here, $m_{\mathrm{p}}$ is the proton mass, and $U_{\mathrm{eff}}^{\mathrm{p}}(\mathbf{r^{\mathrm{p}}}, t)$ and $U_{\mathrm{eff}}^{\mathrm{e}}(\mathbf{r^{\mathrm{e}}}, t)$ are the effective potentials, which are defined in the Supplementary Material.

When the fixed proton basis (FPB) method is used, each quantum proton is associated with a set of fixed electronic and protonic basis functions centered at predetermined coordinates. These fixed basis functions are placed at locations where the proton is expected to move. Expanding the electronic and protonic orbitals in these electronic and protonic basis sets and rearranging the equations leads to the von Neumann equations for the electronic and protonic density matrices.

The RT-NEO-TDDFT approach propagates the electronic and protonic densities using these von Neumann equations, which can be expressed in atomic units as
\begin{equation}
i \frac{\partial }{\partial t} \mathbf{P}^{\mathrm{e}}(t) = [\mathbf{F}^{\mathrm{e}}(t), \mathbf{P}^{\mathrm{e}}(t)]
\label{eq:vonn}
\end{equation}
\begin{equation}
i \frac{\partial }{\partial t} \mathbf{P}^{\mathrm{p}}(t) = [\mathbf{F}^{\mathrm{p}}(t), \mathbf{P}^{\mathrm{p}}(t)]\hspace{0.5em}.
\end{equation}

\noindent Here, $\mathbf{F}^{\mathrm{e}}(t)$ and $\mathbf{F}^{\mathrm{p}}(t)$ are the electronic and protonic Kohn-Sham matrices, and $\mathbf{P}^{\mathrm{e}}(t)$ and $\mathbf{P}^{\mathrm{p}}(t)$ are the electronic and protonic density matrices. These quantities are given in the Supplemental Material. 

The classical nuclei can be propagated using the NEO Ehrenfest approach according to the following equations of motion:
\begin{equation}
M_I \ddot{\mathbf{R}}_I(t) = - \nabla_{\mathbf{R}_I} E\hspace{0.5em}.
\label{eq:grad}
\end{equation}

\noindent Here, $M_I$ is the mass and $\mathbf{R}_I(t)$ is the position of the $I$th classical nucleus. $E$ is the energy of the system, which is a function of the electronic and protonic density matrices, as well as the positions and momenta of the classical nuclei. The explicit form of the energy $E$ is given in section \ref{sec:energy}.

For electronically adiabatic processes, the propagation of the electronic density matrix can be avoided by solving the self-consistent field (SCF) equations to obtain the electronic ground state for the proton density and classical nuclear positions at each time step. This electronic Born-Oppenheimer NEO Ehrenfest dynamics approach allows the use of a much larger time step.\cite{10.1063/5.0142007}

\subsection{RT-NEO-TDDFT with the Original TPB}

The original TPB method has been shown to provide dynamics in agreement with the results of the FPB method.\cite{10.1063/5.0031019,zhao_excited_2021} 
In the original TPB method, the proton orbital is expanded in a set of basis functions with time-dependent centers. The electronic orbitals associated with this time-dependent center also move, but the additional time-dependent terms in the equations of motion are neglected because the electronic basis set is sufficient to describe this movement. The proton orbital is given by
\begin{equation}
\psi^{\mathrm{p}}(\mathbf{r^{\mathrm{p}}}, t) = \sum_k c_k^{\mathrm{p}}(t) \phi_k^{\mathrm{p}}(\mathbf{r^{\mathrm{p}}} - \mathbf{R}^{\mathrm{p}}(t)) \hspace{0.5em}.
\label{eq:Basis}
\end{equation}

\noindent Here, $c_k$ are the coefficients, $\phi_k^{\mathrm{p}}$ are the basis functions, and $\mathbf{R}^{\mathrm{p}}(t)$ is the basis function center. We assume the basis functions have been orthonormalized by a transformation matrix derived from the overlap matrix. In this approach, the movement of the proton basis function center allows the basis functions to move with the proton. 

Substituting Eq. \ref{eq:Basis} into Eq. \ref{eq:TDSE} and multiplying on the left by $\langle\phi_j^{\mathrm{p}}(\mathbf{r^{\mathrm{p}}} - \mathbf{R}^{\mathrm{p}}(t))|$ leads to
\begin{multline}
\langle\phi_j^{\mathrm{p}}(\mathbf{r^{\mathrm{p}}} - \mathbf{R}^{\mathrm{p}}(t))|i \frac{\partial }{\partial t} \sum_k c_k^{\mathrm{p}}(t) |\phi_k^{\mathrm{p}}(\mathbf{r^{\mathrm{p}}} - \mathbf{R}^{\mathrm{p}}(t))\rangle \\ = \langle\phi_j^{\mathrm{p}}(\mathbf{r^{\mathrm{p}}} - \mathbf{R}^{\mathrm{p}}(t))|\hat{F}^{\mathrm{p}}(\mathbf{r^{\mathrm{p}}}, t) \sum_k c_k^{\mathrm{p}}(t) |\phi_k^{\mathrm{p}}(\mathbf{r^{\mathrm{p}}} - \mathbf{R}^{\mathrm{p}}(t))\rangle\hspace{0.5em}.
\label{eq:plug1}
\end{multline}

Using matrix notation, Eq. \ref{eq:plug1} is simplified to
\begin{equation}
i \frac{\partial}{\partial t}\mathbf{C}^{\mathrm{p}} (t) +i \mathbf{\boldsymbol{\tau}} (t) \mathbf{C}^{\mathrm{p}} (t)=  \mathbf{F}^{\mathrm{p}}( t)\mathbf{C}^{\mathrm{p}} (t)
\end{equation}

\noindent where $\mathbf{C}^{\mathrm{p}} (t)$ is the vector of coefficients, and 
\begin{equation}
\mathbf{\boldsymbol{\tau}}_{kj} (t) = \left\langle\phi_k^{\mathrm{p}}(\mathbf{r^{\mathrm{p}}} - \mathbf{R}^{\mathrm{p}}(t))\Bigg| \frac{\partial \phi_j^{\mathrm{p}}(\mathbf{r^{\mathrm{p}}} - \mathbf{R}^{\mathrm{p}}(t))}{\partial \mathbf{R}^{\mathrm{p}}(t)}\right\rangle \cdot\dot{\mathbf{R}}^{\mathrm{p}}(t)\hspace{0.5em}.
\end{equation}
The von Neumann equation for the quantum proton becomes
\begin{equation}
i \frac{\partial }{\partial t} \mathbf{P}^{\mathrm{p}}(t) = [\mathbf{F}^{\mathrm{p}}(t), \mathbf{P}^{\mathrm{p}}(t)] - i\left( \boldsymbol{\tau} (t) \mathbf{P}^{\mathrm{p}}(t) +\mathbf{P}^{\mathrm{p}}(t)\boldsymbol{\tau}^{\dagger}(t) \right)\hspace{0.5em}.
\label{eq:otpb}
\end{equation}
%

In the original TPB method, the protonic basis function center is treated classically and propagated according to Eq. \ref{eq:grad}, where the fictitious mass is the proton mass. The protonic density matrix is propagated using Eq. \ref{eq:otpb}, and the electronic density matrix is propagated using Eq. \ref{eq:vonn}, except when the electronic Born--Oppenheimer approximation is invoked. In the original TPB method, the system energy is defined without the kinetic energy associated with $\dot{\mathbf{R}}^{\mathrm{p}}(t)$ and is conserved in the complete basis set limit. 
In practice, however, RT-NEO-TDDFT Ehrenfest dynamics calculations use a finite basis set. Previous work found the energy to be poorly conserved for proton transfer reactions, although the dynamics are in good agreement with the benchmark FPB calculations.\cite{zhao_excited_2021} A TPB method that rigorously conserves energy as well as providing accurate dynamics is desirable.

The semiclassical TPB (sc-TPB) method used an extended Lagrangian  formalism to address these issues to some extent.\cite{10.1063/5.0230570} This method resulted in improved energy conservation with slightly altered dynamics compared to the benchmark FPB method. Our c-TPB approach starts out with the same wavefunction ansatz but retains a term that was neglected in the sc-TPB method and applies a constraint that results in  more stable dynamics.

\subsection{Constrained traveling proton basis RT-NEO-TDDFT}

To derive the c-TPB approach, we start with the following wavefunction ansatz:
\begin{equation}
\psi^{\mathrm{p}}(\mathbf{r^{\mathrm{p}}}, t) = \sum_k c_k^{\mathrm{p}}(t)  e^{i m_{\mathrm{p}} \dot{\mathbf{R}}^{\mathrm{p}}(t) \cdot \mathbf{r^{\mathrm{p}}}}\phi_k^{\mathrm{p}}(\mathbf{r^{\mathrm{p}}} - \mathbf{R}^{\mathrm{p}}(t))
\label{eq:Basis2}
\end{equation}
\noindent where $m_{\mathrm{p}}$ is the mass of the proton. This phase factor could be viewed as a momentum boost or a quantum translation. Substituting Eq. \ref{eq:Basis2} into Eq. \ref{eq:TDSE} and multiplying on the left by $\langle\phi_j^{\mathrm{p}}(\mathbf{r^{\mathrm{p}}} - \mathbf{R}^{\mathrm{p}}(t)) |e^{-i m_{\mathrm{p}} \dot{\mathbf{R}}^{\mathrm{p}}(t) \cdot \mathbf{r^{\mathrm{p}}}}$ leads to
\begin{multline}
\langle\phi_j^{\mathrm{p}}(\mathbf{r^{\mathrm{p}}} - \mathbf{R}^{\mathrm{p}}(t))|e^{-i m_{\mathrm{p}} \dot{\mathbf{R}}^{\mathrm{p}}(t) \cdot \mathbf{r^{\mathrm{p}}}} i \frac{\partial }{\partial t} \sum_k c_k^{\mathrm{p}}(t) e^{i m_{\mathrm{p}} \dot{\mathbf{R}}^{\mathrm{p}}(t) \cdot \mathbf{r^{\mathrm{p}}}} \\ |\phi_k^{\mathrm{p}}(\mathbf{r^{\mathrm{p}}} - \mathbf{R}^{\mathrm{p}}(t))\rangle = \langle\phi_j^{\mathrm{p}}(\mathbf{r^{\mathrm{p}}} - \mathbf{R}^{\mathrm{p}}(t))|  e^{-i m_{\mathrm{p}} \dot{\mathbf{R}}^{\mathrm{p}}(t) \cdot \mathbf{r^{\mathrm{p}}}} \hat{F}^{\mathrm{p}}(\mathbf{r^{\mathrm{p}}}, t)\\ \sum_k c_k^{\mathrm{p}}(t) e^{i m_{\mathrm{p}} \dot{\mathbf{R}}^{\mathrm{p}}(t) \cdot \mathbf{r^{\mathrm{p}}}}|\phi_k^{\mathrm{p}}(\mathbf{r^{\mathrm{p}}} - \mathbf{R}^{\mathrm{p}}(t))\rangle\hspace{0.5em}.
\label{eq:plug}
\end{multline}

Computing the derivatives and converting to matrix notation, this equation is simplified to
\begin{multline}
i \frac{\partial}{\partial t}\mathbf{C}^{\mathrm{p}} (t) +i \mathbf{\tau} (t) \mathbf{C}^{\mathrm{p}} (t) - \mathbf{A}(t)\mathbf{C}^{\mathrm{p}} (t) \\=  \mathbf{F}^{\mathrm{p}}( t)\mathbf{C}^{\mathrm{p}} (t)+\frac{1}{2} m_{\mathrm{p}} (\dot{\mathbf{R}}^{\mathrm{p}}(t))^2 \mathbf{C}^{\mathrm{p}} (t)+i \mathbf{\tau} (t) \mathbf{C}^{\mathrm{p}} (t)
\label{eq:ndiff}
\end{multline}
\noindent where $\mathbf{A}(t)$ has matrix elements
\begin{equation}
\mathbf{A}_{kj}(t) =  m_{\mathrm{p}}\langle\phi_k^{\mathrm{p}}(\mathbf{r^{\mathrm{p}}} - \mathbf{R}^{\mathrm{p}}(t))|\mathbf{r^{\mathrm{p}}} | \phi_j^{\mathrm{p}}(\mathbf{r^{\mathrm{p}}} - \mathbf{R}^{\mathrm{p}}(t))\rangle \cdot \ddot{\mathbf{R}}^{\mathrm{p}}(t)\hspace{0.5em}.
\label{eq:Amat}
\end{equation}

The von Neumann equation obtained from Eq. \ref{eq:ndiff} is
\begin{equation}
i \frac{\partial }{\partial t} \mathbf{P}^{\mathrm{p}}(t) = [\mathbf{F}^{\mathrm{p}}(t)+\mathbf{A}(t), \mathbf{P}^{\mathrm{p}}(t)] 
\label{eq:ntpb}
\end{equation}

\noindent where the $\frac{1}{2} m_{\mathrm{p}} (\dot{\mathbf{R}}^{\mathrm{p}}(t))^2$ term is not included because it is a spatial constant at any given time. Eq. \ref{eq:ntpb} is analytically exact for any choice of $\ddot{\mathbf{R}}^{\mathrm{p}}(t)$ within the complete basis set limit. 
Note that the electronic von Neumann equation will also in general contain additional terms due to the TPB, but these terms are negligible because the electronic basis set is sufficient to describe electronic fluctuations due to the much lighter mass of electrons.

Eq. \ref{eq:ntpb} is the same as Eq. 23 in Ref. ~\onlinecite{10.1063/5.0230570}, which provides the derivation of the sc-TPB method. That previous derivation neglects the $\mathbf{A}(t)$ term due to numerical instabilities in the sc-TPB method. \cite{10.1063/5.0230570} \nick{Our c-TPB method solves this issue by applying a constraint, as discussed in the next subsection.}

\subsubsection{Constraining the proton basis function center}

The RT-NEO-TDDFT c-TPB Ehrenfest dynamics method is derived using an extended Lagrangian formulation in the Supplementary Material. The equations of motion for the classical nuclei are given by Eq. \ref{eq:grad}. The analogous equations of motion are used for the protonic basis function center:
\begin{equation}
m_{\mathrm{p}}\ddot{ \mathbf{R}}^{\mathrm{p}}(t) = -\nabla_{\mathbf{R}^{\mathrm{p}} } E\hspace{0.5em}.
\label{eq:grad2}
\end{equation}

For the quantum subsystem, Eq. \ref{eq:vonn} is used for propagating the electronic density, and Eq. \ref{eq:ntpb} is used for propagating the protonic density. However, to ensure that the protonic basis function center follows the proton motion, we 
apply a constraint such that ${\mathbf{R}}^{\mathrm{p}}(t) = \langle \psi^{\mathrm{p}}(\mathbf{r^{\mathrm{p}}}, t)|\mathbf{r^{\mathrm{p}}}| \psi^{\mathrm{p}}(\mathbf{r^{\mathrm{p}}}, t)\rangle \equiv \langle \mathbf{r^{\mathrm{p}}} \rangle(t)$ at all times $t$. For this purpose, we employ a constrained dynamics approach, and Eq. \ref{eq:Amat} is modified to 
\begin{equation}
\mathbf{A}_{kj}(t) =  m_{\mathrm{p}}\langle\phi_k^{\mathrm{p}}(\mathbf{r^{\mathrm{p}}} - \mathbf{R}^{\mathrm{p}}(t))|\mathbf{r^{\mathrm{p}}} | \phi_j^{\mathrm{p}}(\mathbf{r^{\mathrm{p}}} - \mathbf{R}^{\mathrm{p}}(t))\rangle \cdot \mathbf{f}(t)\hspace{0.5em}.
\label{eq:lagrange}
\end{equation}

\noindent Here, $\mathbf{f}(t)$ is a Lagrange multiplier that is optimized at each time step to ensure that 
${\mathbf{R}}^{\mathrm{p}}(t) = \langle \mathbf{r^{\mathrm{p}}} \rangle(t)$.
Specifically, the Lagrange multiplier is determined by solving the following equation: 
\begin{equation}
\mathbf{f}^{a+1}(t + \Delta t) =\mathbf{f}^{a}(t+ \Delta t)  -\frac{2 }{\Delta t^2} (\mathbf{R}^{\mathrm{p}}(t+ \Delta t) -\langle  \mathbf r^{\mathrm{p}}  \rangle^{a}(t+ \Delta t) )
\label{eq:ulm}
\end{equation}

\noindent
where $\mathbf{f}^a(t)$ is the Lagrange multiplier at the $a$th iteration. At $t=0$, $\mathbf{f}^1(0)=0$, and at $t= m\Delta t$, $\mathbf{f}^1(m\Delta t)=\mathbf{f}((m-1)\Delta t)$, where $m$ is an integer. The Lagrange multiplier is optimized until 
\begin{equation}
|\langle \mathbf{r^{\mathrm{p}}} \rangle(t+\Delta t) -\mathbf{R}^{\mathrm{p}}(t+\Delta t) | < \epsilon 
\label{eq:lm}
\end{equation}
\noindent
where $\epsilon$ is a specified threshold value.
This constrained Lagrange multiplier NEO real-time dynamics approach leads to accurate and stable dynamics.

Note that this constrained approach is different from the constrained NEO \nick{(CNEO)} approach developed by Yang and coworkers, which constrains the proton density \nick{during the self-consistent field} solution of the time-independent nuclear--electronic Schr\"odinger equation.\cite{10.1063/1.5143371} \nick{Both approaches apply a constraint to the expectation value of the proton position operator.} However, our approach constrains the proton density in a {\it nonequilibrium} state during the propagation of the {\it time-dependent} nuclear--electronic Schr\"odinger equation for real-time dynamics. \nick{Our constraint alters the time-dependent von Neumann equations in Eq. \ref{eq:ntpb} by substituting Eq. \ref{eq:lagrange} for $\mathbf{A}_{kj}(t)$. Both approaches can be used to compute vibrational spectra. The CNEO approach performs classical molecular dynamics for all nuclei on the constrained minimum energy surface generated by solution of the time-independent nuclear--electronic Schr\"odinger equation. In contrast, the RT-NEO-TDDFT c-TPB Ehrenfest dynamics approach integrates the time-dependent nuclear--electronic Schr\"odinger equation to propagate the protonic densities, quenching the electronic density at each time step for simulating purely vibrational spectra but propagating the electronic densities in real time for other situations. Thus, the RT-NEO-TDDFT c-TPB Ehrenfest dynamics approach can also be used to compute vibronic spectra (i.e., including electronic, vibrational, and vibronic excitations), as well as to simulate the nonequilibrium, real-time dynamics of the electronic and protonic densities, in conjunction with the other nuclear degrees of freedom.}

\subsubsection{\label{sec:energy}Energy within the c-TPB approach}

Within the c-TPB approach, the total energy of the system is given by 

\begin{multline}
\label{eq:energy}
E  =  \text{Tr} \left[\right. \mathbf{H}^{\mathrm{e}}_{\text{core}} \mathbf{P}^{\mathrm{e}} \left]\right. + \text{Tr} \left[\right. \mathbf{H}^{\mathrm{p}}_{\text{core}} \mathbf{P}^{\mathrm{p}} \left]\right. + \frac{1}{2} \text{Tr} \left[\right. \mathbf{J}^{\mathrm{ee}} \mathbf{P}^{\mathrm{e}} \left]\right. 
\\+ \textrm{ } \frac{1}{2} \text{Tr} \left[\right. \mathbf{J}^{\mathrm{pp}} \mathbf{P}^{\mathrm{p}} \left]\right. 
- \text{Tr} \left[\right. \mathbf{J}^{\mathrm{ep}} \mathbf{P}^{\mathrm{p}} \left]\right. + \textrm{ }  E_{\mathrm{exc}} \left[\right. \mathbf{P}^{\mathrm{e}} \left]\right. + E_{\mathrm{pxc}} \left[\right. \mathbf{P}^{\mathrm{p}} \left]\right.\\ + E_{\mathrm{epc}} \left[\right. \mathbf{P}^{\mathrm{e}} , \mathbf{P}^{\mathrm{p}} \left]\right. + V_{\mathrm{NN}} + \sum_I^{N_{\mathrm c}} \frac{1}{2} M_I \dot{\mathbf{ R}}_I^2 + \sum_n^{N_{\rm p}} \frac{1}{2} m_{\mathrm{p}} (\dot{\mathbf{ R}}^{\mathrm{p}}_n)^2\hspace{0.5em}.
\end{multline}

\noindent Here, $\mathbf{H}_{\text{core}}^{\mathrm{e/p}}$  are the electron/proton one-particle core matrices, $\mathbf{J}^{\mathrm{e e/pp}}$ are the electron--electron/proton--proton Coulomb interaction matrices, and $\mathbf{J}^{\mathrm{ep}}$ is the electron--proton Coulomb interaction matrix. Moreover, $E_{\mathrm{exc}}$ is the electron exchange--correlation energy, $E_{\mathrm{pxc}}$ is the proton exchange--correlation energy, and $E_{\mathrm{epc}}$ is the electron--proton correlation energy. $V_{\mathrm{NN}}$ is the repulsive Coulomb interaction energy between classical nuclei, $\sum_I^{N_{\mathrm c}} \frac{1}{2} M_I \dot {{\mathbf{R}}}_I^2$ is the kinetic energy of the classical nuclei, and $\sum_n^{N_{\rm p}} \frac{1}{2} m_{\mathrm{p}} (\dot{\mathbf{ R}}^{\mathrm{p}}_n)^2$ is the kinetic energy of the protonic basis function centers. 

\subsection{Multiple Quantum Protons and Propagation Scheme}

In the case of multiple quantum protons, each proton is treated as a distinguishable particle using the nuclear Hartree product representation. \cite{auer_localized_2010, 10.1063/5.0308634} As shown in in the Supplementary Material, the key equations for multiple quantum protons become
\begin{equation}
m_{\mathrm{p}} \ddot{ \mathbf{R}}_n^{\mathrm{p}}(t)  =   -\nabla_{\mathbf{R}_n^{\mathrm{p}} } E 
\label{eq:gradients}
\end{equation}
\begin{equation}
i \frac{\partial }{\partial t} \mathbf{P}_n^{\mathrm{p}}(t) = [\mathbf{F}_n^{\mathrm{p}}(t) + \mathbf{A}_n(t), \mathbf{P}_n^{\mathrm{p}}(t)]
\label{eq:hptd}
\end{equation}
\noindent 
\begin{equation}
\begin{aligned}
(\mathbf{A}_n)_{kj}(t) =&m_{\mathrm{p}} \langle\phi_{n, k}^{\mathrm{p}}(\mathbf{r^{\mathrm{p}}} - \mathbf{R}^{\mathrm{p}}_{n}(t))|\mathbf{r^{\mathrm{p}}} | \phi_{n, j}^{\mathrm{p}}(\mathbf{r^{\mathrm{p}}} - \mathbf{R}_{n}^{\mathrm{p}}(t))\rangle \\ &\cdot \mathbf{{f}}_{n}(t) 
\label{eq:amat2}
\end{aligned}
\end{equation}
\begin{equation}
\begin{aligned}
\mathbf{f}_{n}^{a+1}(t + \Delta t) =&\mathbf{f}_{n}^{a}(t+ \Delta t)  \\ &-\frac{2 }{\Delta t^2} (\mathbf{R}_n^{\mathrm{p}}(t+ \Delta t) -\langle  \mathbf r^{\mathrm{p}} \rangle^a_n(t + \Delta t) )
\end{aligned}
\label{eq:lm2}
\end{equation}
\noindent
where $ \mathbf{R}_n^{\mathrm{p}}$ is the $n$-th protonic basis function center, $\mathbf{P}_n^{\mathrm{p}}$ is the $n$-th protonic density matrix, $\mathbf{F}_n^{\mathrm{p}}$ is the $n$-th protonic Kohn-Sham matrix,  $\mathbf{A}_n$ is the $n$-th protonic $\mathbf A$ matrix, $\mathbf{f}_{n}$ is the $n$-th protonic Lagrange multiplier, $\phi_{n, j}^{\mathrm{p}}$ are the $n$-th protonic basis functions, and $\langle  \mathbf r^{\mathrm{p}} \rangle^a_n$ is the  $n$-th proton position expectation value at the $a$-th iteration. 

The c-TPB algorithm is shown in Algorithm 1.
In this algorithm, $N_{\text{steps}}$ is the total number of time steps, $\mathbf{R}_{I}(t)$ are the classical nuclear coordinates, $\mathbf{R}_{n}^{\text{p}}(t)$ are the traveling protonic basis function centers, $N_{\text{nuc}}$ is the number of steps propagating the electronic and protonic density matrices between gradient calculations, $\Delta t_{\rm q}$ is the time step for the density propagation, $\Delta t = N_{\text{nuc}} \Delta t_{\rm q}$ is the time step for the propagation of the classical nuclei and protonic basis function centers, and $N_{\text{p}}$ is the number of protons. The leapfrog equations are given by
\begin{equation}
\dot {\mathbf{R}}_I \left(t+\frac{\Delta t}{2}\right) = \dot {\mathbf{R}}_I \left(t-\frac{\Delta t}{2}\right) + \ddot {\mathbf{R}}_I (t) \Delta t,
\label{eq:vv1}
\end{equation}
\begin{equation}
 {\mathbf{R}}_I (t+\Delta t) = {\mathbf{R}}_I (t) +\dot {\mathbf{R}}_I \left(t+\frac{\Delta t}{2}\right)  \Delta t \hspace{0.5em}.
\label{eq:vv2}
\end{equation}
\noindent
Similarly, for the traveling protonic basis function centers, the leapfrog equations are
\begin{equation}
\dot {\mathbf{R}}_n^{\text{p}} \left(t+\frac{\Delta t}{2}\right) = \dot {\mathbf{R}}_n^{\text{p}} \left(t-\frac{\Delta t}{2}\right) + \ddot {\mathbf{R}}_n^{\text{p}} (t) \Delta t ,
\label{eq:vv1q}
\end{equation}
\begin{equation}
 {\mathbf{R}}_n^{\text{p}} (t+\Delta t) = {\mathbf{R}}_n^{\text{p}} (t) +\dot {\mathbf{R}}_n^{\text{p}} \left(t+\frac{\Delta t}{2}\right)  \Delta t \hspace{0.5em}.
\label{eq:vv2q}
\end{equation}



\begin{figure*}
\centering
    \includegraphics[width=\textwidth]{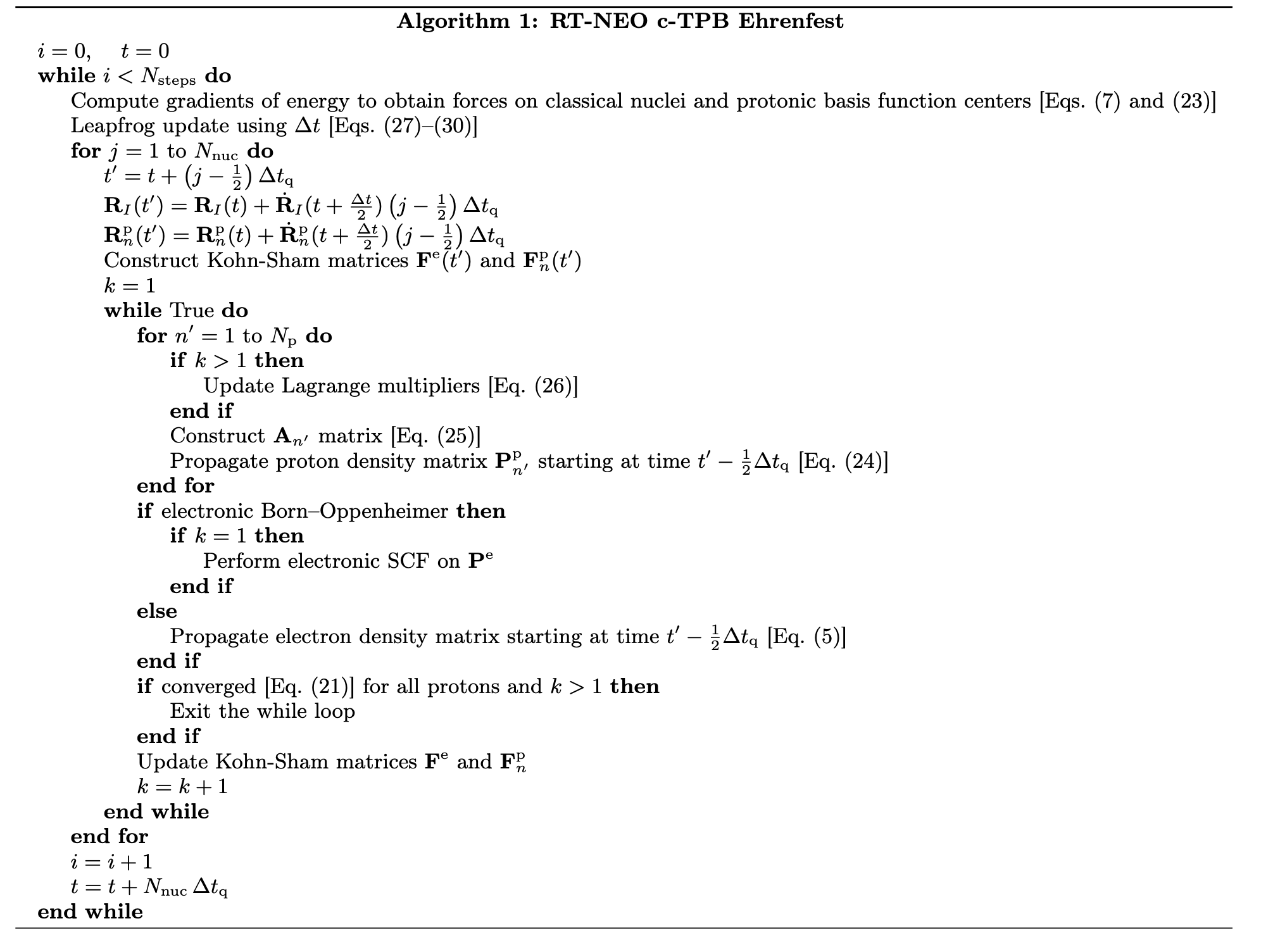}
\end{figure*}

\section{Simulation Details}

In this section, we provide the simulation parameters and computational details for the vibrational analyses and proton transfer dynamics presented in Section IV. The geometries are given in the Supplementary Material. All simulations were performed with our implementation of the  c-TPB method in a development branch of the Q-Chem software. \cite{10.1063/5.0055522} 

\subsection{Vibrational frequencies}

As an initial test of the c-TPB method, we simulated the vibrational frequencies of molecules containing single and multiple quantum protons. The single quantum proton molecules studied are HCN, HNC, and FHF$^-$, and the multiple quantum proton molecules studied are H$_2$O, HCOOH, H$_2$CO, and H$_2$. The B3LYP electronic exchange--correlation functional\cite{becke_densityfunctional_1993} and the epc17-2 electron-proton correlation functional\cite{yang_development_2017, brorsen_multicomponent_2017} were used. The cc-pV5Z electronic basis set\cite{dunning_gaussian_1989} and PB4-D protonic basis set\cite{yu_development_2020} were used for the hydrogen atoms, and the cc-pVDZ electronic basis set\cite{dunning_gaussian_1989} was used for all other atoms. The time step, $\Delta t_{\rm q}$, was 4 a.u. for all vibrational frequency trajectories. \nick{Each trajectory was propagated for 4000 steps for a total of $\sim $ 387 fs, with the exception of the original TPB trajectories with multiple protons, which were propagated for half this time because some of them became numerically unstable at longer times. The original TPB trajectory for FHF$^-$ also became numerically unstable at longer times, and therefore only the first 97 fs of the trajectory was used to compute the frequencies. The c-TPB trajectories were numerically stable in all cases studied and shown to be converged. For all calculations, $N_{\rm nuc} = 1$ and $\epsilon = 10^{-4}$. The c-TPB trajectories were initiated with geometries optimized using CNEO-DFT with the nuclear Hartree product representation. The original TPB trajectories were initiated with geometries optimized using NEO-DFT with the nuclear Slater determinant representation.}

The electronic Born--
Oppenheimer NEO Ehrenfest dynamics method\cite{10.1063/5.0142007} was used for all vibrational spectra simulations. To initially perturb the system, a  small displacement of $1 \times 10^{-5}$ a.u. was applied to all three Cartesian coordinates of a classical nucleus (or a quantum nucleus for the case of H$_2$) at the optimized geometry. The perturbed atom is the first atom listed in each molecule in the geometries given in the Supplementary Material. The velocities were initially set to zero. The frequencies were obtained by a Fourier transform of the time-dependent total dipole moment. We also performed VPT2 calculations\cite{10.1063/1.1824881} with the same electronic basis sets and functional to provide a benchmark for the RT-NEO calculations. 

\subsection{Excited-state intramolecular proton transfer reactions}
\label{sec:esipt}
\begin{figure}
    \centering
    \includegraphics[width=\linewidth]{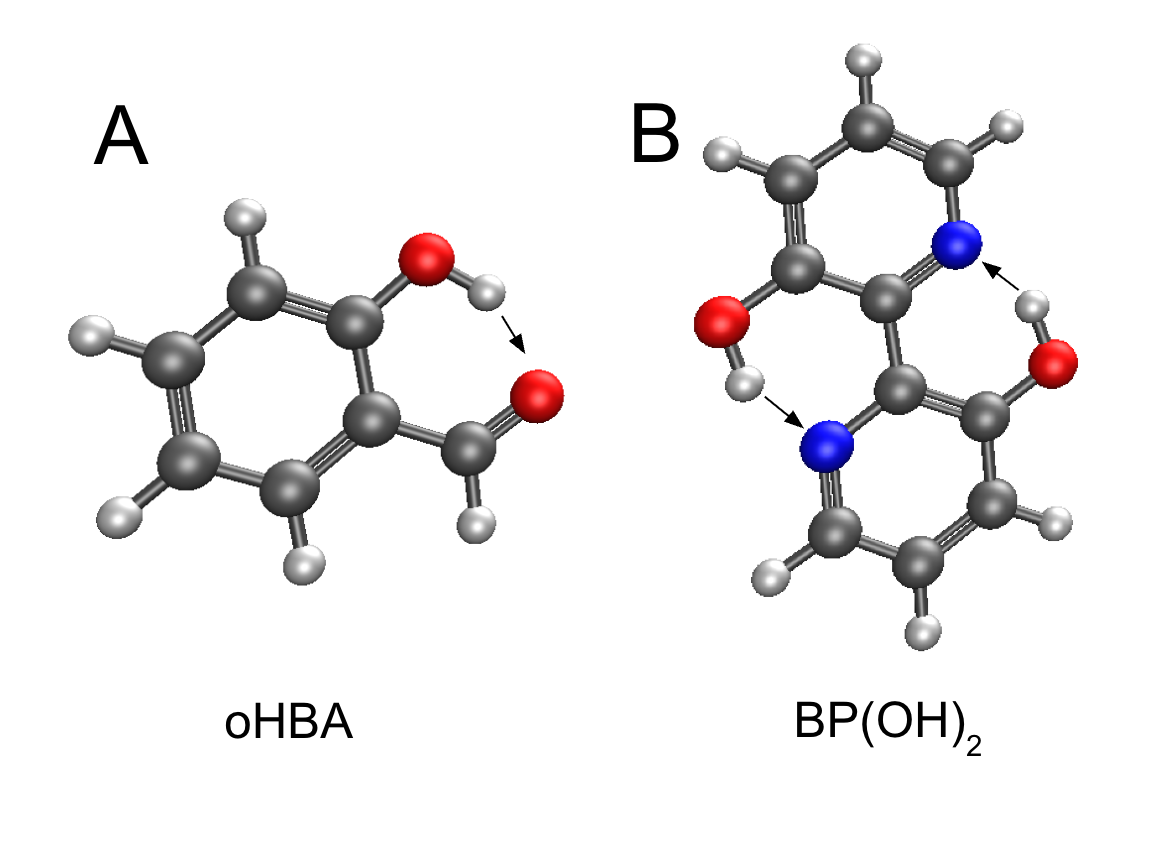}
    \caption{(A) The oHBA geometry with an arrow showing the proton transfer. (B) The BP(OH)$_2$ geometry with two arrows showing the double proton transfer.}
    \label{fig:moles}
\end{figure}

As another illustrative example, we simulated excited-state intramolecular proton transfer in the \textit{o}-hydroxybenzaldehyde (oHBA) molecule and double proton transfer in the [2,2'-bipyridyl]-3,3'-diol (BP(OH)$_2$) molecule. Figure \ref{fig:moles} shows both molecules and the transferring protons. Again, the B3LYP electronic exchange--correlation functional and the epc17-2 electron-proton correlation functional were used.  The cc-pVDZ electronic basis set and PB4-F2 protonic basis set\cite{yu_development_2020} were used for oHBA, and the 6-31G electronic basis set\cite{hehre_selfconsistent_1972} and PB4-D protonic basis set were used for BP(OH)$_2$. \nick{As for the vibrational frequency calculations, the c-TPB trajectories were initiated with geometries optimized using CNEO-DFT, and the original TPB trajectories were initiated with geometries optimized using NEO-DFT.}

Both systems were initialized with an electronic HOMO to LUMO excitation. The RT-NEO-TDDFT Ehrenfest dynamics were propagated with gradients computed every 10 steps, i.e., $N_{\rm nuc} =10$, and the threshold for the constraint(s) was $\epsilon = 10^{-4}$.
For the oHBA molecule, we performed RT-NEO-TDDFT Ehrenfest dynamics with the original TPB method, \cite{10.1063/5.0031019} the c-TPB method, and the FPB method. The FPB method used four fixed protonic basis function centers placed along the trajectory obtained with the c-TPB method for the first 18.5 fs. For the TPB oHBA calculations, the timestep, $\Delta t_{\rm q}$, was  0.04 a.u., and for the FPB oHBA calculation, the timestep was  0.4 a.u. due to the greater computational expense. For all oHBA calculations, the trajectories were propagated for $\sim 29$ fs. 
For the BP(OH)$_2$ molecule, we only performed RT-NEO-TDDFT Ehrenfest dynamics with the c-TPB method. In this case, the  timestep was 0.1 a.u., and the trajectory was propagated for \nick{18} fs.

\section{Results}

\subsection{Vibrational frequencies}

The vibrational frequencies for the single proton and multiple proton systems are given in Table \ref{tab:hcn} and Table \ref{tab:multiproton}, respectively. 
The VPT2 vibrational frequencies computed at the same level of electronic structure theory are given as a reference. The mean absolute error (MAE) for the vibrational modes of each system using this reference is also given. Overall, the MAE is lower for frequencies obtained with the c-TPB method than for frequencies obtained with the original TPB method. 
For the few individual modes exhibiting slightly smaller errors with the original TPB method, the original and c-TPB methods produce very similar vibrational frequencies. Most importantly, the errors for most of the stretching modes involving protons are significantly smaller when computed with the c-TPB method compared to the original TPB method. \nick{The frequencies were shown to be converged by comparing results from shorter and longer trajectories (Tables S1 and S2). Full convergence of the frequencies obtained with the original TPB method is challenging due to numerical instabilities that arise for longer trajectories. A significant advantage of the c-TPB method is the substantially improved numerical stability.} These results demonstrate the ability of the c-TPB method to produce accurate vibrational spectra.

\begin{table*}[ht]
\centering
\begin{threeparttable}

\caption{HCN, HNC, and FHF$^-$ vibrational frequencies (cm$^{-1}$) with absolute errors relative to the VPT2 reference given in parentheses.}
\label{tab:hcn}
\begin{tabular}{llccc}
\hline
Molecule & Vibrational mode & Ref.\ VPT2 & Original TPB & c-TPB \\
\hline
HCN
& CH bend    & 760  & 808 (48)    & 762 (2) \\
& CN stretch & 2176 & 2242 (66)   & 2196 (20) \\
& CH stretch & 3328 & 3493 (165)  & 3345 (17) \\
& MAE        & ---  & 93 & 13 \\
\hline
HNC
& NH bend    & 457  & 465 (8)     & 444 (13) \\
& NC stretch & 2068 & 2115 (47)   & 2099 (31) \\
& NH stretch & 3626 & 3912 (287)  & 3604 (21) \\
& MAE        & ---  & 114 & 22 \\
\hline
FHF$^-$\tnote{a}
& FF stretch & 592  & 636 (43)    & 638 (46) \\
& FH bend    & 1351 & 1387 (36)   & 1354 (3) \\
& FH stretch & 1721 & 2080 (359)   & 2006 (286) \\
& MAE        & ---  & 146 & 111 \\
\hline
\end{tabular}
\begin{tablenotes}
\footnotesize
\item[a] \nick{Only data from the first 97 fs of the original TPB trajectory for FHF$^-$ was used due to numerical instabilities occurring at later times.}
\end{tablenotes}
\end{threeparttable}
\end{table*}

Within the NEO framework, molecular rotations are not constrained unless at least two nuclei are treated classically. This issue leads to a small amount of rotational contamination of the molecular vibrational zero-point energies, defined as the difference between the NEO-DFT and conventional DFT energies, each at their respective optimized structure. This rotational contamination of the total vibrational zero-point energy (ZPE) also arises for the constrained NEO method,\cite{10.1063/1.5143371} where constraints are applied to the position expectation values of the quantum protons, as shown for an isolated water molecule.\cite{doi:10.1021/acs.jpclett.5c03144} However, when the frequencies are computed from constrained dynamics, namely the Fourier transfer of the time-dependent dipole moment, the rotational contamination becomes almost negligible, as shown in Table \ref{tab:h2o_zpe}.

This characteristic is evident in our simulations of the H$_2$ and H$_2$O molecules.  For the original TPB method, the RT-NEO-TDDFT Ehrenfest dynamics simulations of H$_2$ and H$_2$O were numerically unstable because the protonic basis function centers became distinct from the corresponding proton position expectation values. We also found that the previously developed sc-TPB method,\cite{10.1063/5.0230570} which does not include the ${\bf A}$ matrix or the Lagrangian constraints, did not produce an accurate vibrational frequency for H$_2$. However, the c-TPB method was numerically stable for the H$_2$ and H$_2$O molecules and produced accurate vibrational frequencies, as shown in Table \ref{tab:multiproton}.  

These results illustrate the capability of the c-TPB method to more accurately capture the vibrational frequencies compared to the original TPB method, particularly for cases where rotational contamination may be significant. In every molecule studied, the c-TPB method produced a lower MAE compared to the original TPB method.

\begin{table*}[t]
\centering
\begin{threeparttable}

\caption{
H$_2$O, HCOOH, H$_2$CO, and H$_2$ vibrational frequencies (cm$^{-1}$) with absolute errors relative to the VPT2 reference given in parentheses.
}
\label{tab:multiproton}
\begin{tabular}{llccc}
\hline
Molecule & Vibrational Mode & Ref.\ VPT2 & Original TPB & c-TPB \\
\hline

{H$_2$O\tnote{a}}
& bending                & 1604 & --- & 1582 (22) \\
& symmetric stretch      & 3644 & --- & 3697 (53) \\
& asymmetric stretch     & 3697 & --- & 3697 (0) \\
&                         & MAE  & --- & 25 \\
\hline

{HCOOH}
& OCO bend               & 626  & 637 (12)  & 622 (3) \\
& torsion                & 638  & 637 (1)   & 632 (6) \\
& CH bend                & 1034 & 1168 (134) & 1043 (8) \\
& CO stretch             & 1111 & 1168 (56) & 1107 (5) \\
& OH bend                & 1253 & 1221 (32) & 1270 (17) \\
& CH bend                & 1394 & 1221 (173) & 1458 (64) \\
& C=O stretch            & 1802 & 1820 (17) & 1820 (18) \\
& CH stretch             & 2906 & 2780 (126) & 2986 (79) \\
& OH stretch             & 3509 & 3076 (433) & 3500 (9) \\
&                         & MAE  & 109 & 23 \\
\hline

{H$_2$CO}
& CH$_2$ wag             & 1188 & 945 (243) & 1196 (8) \\
& CH$_2$ rock            & 1247 & 945 (302) & 1237 (11) \\
& CH$_2$ scissor         & 1517 & 1806 (289) & 1483 (35) \\
& CO stretch             & 1798 & 1806 (8) & 1808 (10) \\
& symmetric CH stretch   & 2739 & 2592 (147) & 2813 (74) \\
& asymmetric CH stretch  & 2749 & 2592 (157) & 2813 (64) \\
&                         & MAE  & 191 & 34 \\
\hline

{H$_2$\tnote{a}}
& stretching             & 4114 & --- & 4040 (74) \\
&                         & MAE  & --- & 74 \\
\hline

\end{tabular}
\begin{tablenotes}
\footnotesize
\item[a] \nick{The original TPB method was numerically unstable for the H$_2$ and H$_2$O systems.}
\end{tablenotes}
\end{threeparttable}
\end{table*}

\begin{table}[t]
\centering
\caption{
Comparison of vibrational frequencies and ZPE
for H$_2$O computed with the c-TPB method against NIST reference data. \cite{shimanouchi_tables_1972}
}
\label{tab:h2o_zpe}
\begin{tabular}{lccc}
\hline
Mode & c-TPB ($\mathrm{cm^{-1}}$) & NIST ($\mathrm{cm^{-1}}$) & Error (\%) \\
\hline
Bending              & 1582 & 1595 & -0.82 \\
Symmetric stretch    & 3697 & 3657 &  1.09 \\
Asymmetric stretch   & 3697 & 3756 & -1.57 \\
\hline
ZPE                  & 4488 & 4504 & -0.36 \\
\hline
\end{tabular}
\end{table}

\subsection{Excited-state intramolecular proton transfer dynamics}

\begin{figure}
    \centering
    \includegraphics[width=\linewidth]{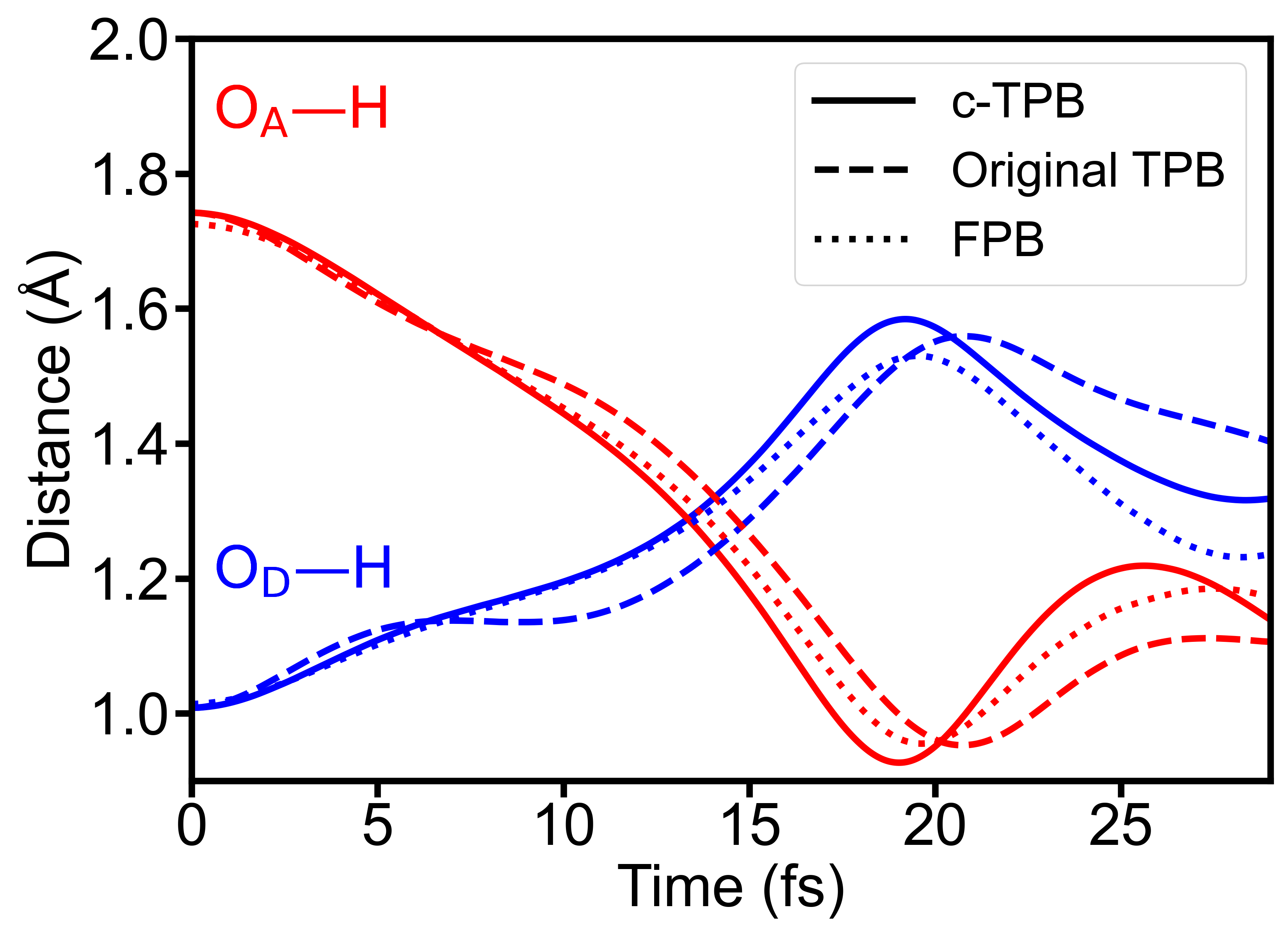}
    \caption{Distance between the transferring proton position expectation value and the donor or acceptor oxygen atom along the trajectories for excited-state intramolecular proton transfer in oHBA, as obtained from RT-NEO-TDDFT Ehrenfest dynamics with the c-TPB, original TPB, and FPB methods. The c-TPB method agrees better with the FPB method, which serves as a reference up to $\sim$ 18 fs.}
    \label{fig:ohba}
\end{figure}

In this section, we investigate excited-state intramolecular proton transfer dynamics. First we study the photoinduced single proton transfer reaction in the oHBA molecule. Figure \ref{fig:ohba} shows the distance between the transferring proton and the donor or acceptor oxygen atom along the trajectories generated with the FPB, original TPB, and c-TPB methods. These distances are calculated using the expectation value of the position operator of the quantum proton. The FPB trajectory is considered to be the reference for early times because no moving basis function centers are necessary. As discussed in Section \ref{sec:esipt}, the protonic basis function centers for the FPB method were placed along the c-TPB trajectory using the first 18.5 fs of that trajectory. The c-TPB and FPB trajectories are in excellent agreement over this time period but then diverge slightly, mainly because the fixed basis function centers in the FPB method are no longer adequate (\nick{Figure S2}). However,the excellent agreement for the portion of the trajectory where the FPB method can be viewed as a reliable reference indicates that the c-TPB method produces accurate proton transfer dynamics. The original TPB method also produces qualitatively reasonable proton transfer dynamics but is clearly not as accurate as the c-TPB method, also exhibiting small unphysical oscillations.

\begin{figure}
    \centering
    \includegraphics[width=\linewidth]{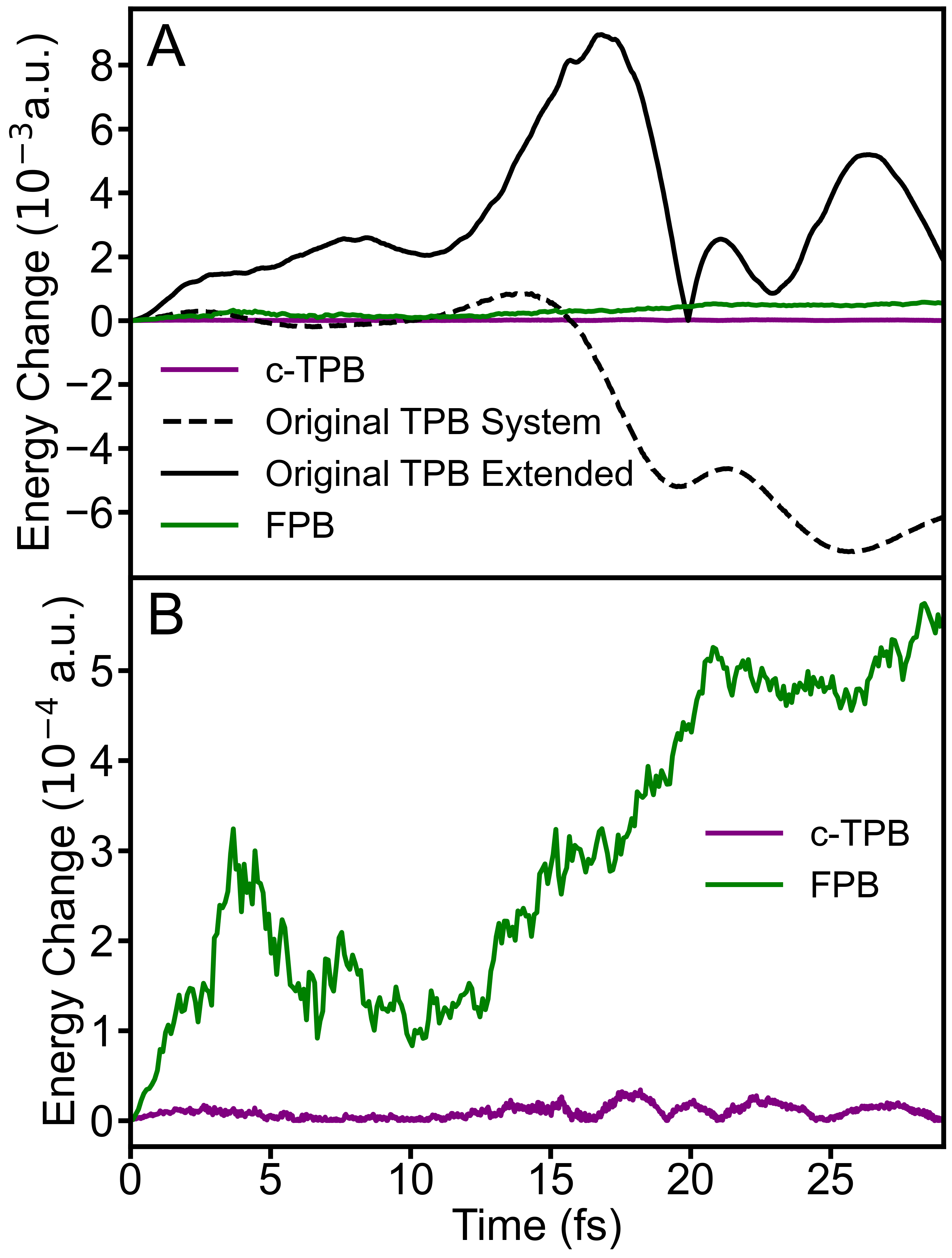}
    \caption{ (A)Energy change along the trajectories for excited-state intramolecular proton transfer in oHBA, as obtained from RT-NEO-TDDFT Ehrenfest dynamics trajectories with the c-TPB (purple), original TPB (black), and FPB (green) methods. For the original TPB method, the system energy and extended energy are shown as dashed and solid lines, respectively, and are not conserved properties. (B) Magnified view of the energy change for the trajectories propagated with the c-TPB and FPB methods. Note that the energy scales for parts A and B differ by two orders of magnitude. }
    \label{fig:ohbae}
\end{figure}

Figure \ref{fig:ohbae} shows the energy change along the trajectory generated with each method. For the c-TPB method, the total energy given by Eq. \ref{eq:energy} was computed as the expectation value of the NEO Hamiltonian using the c-TPB wavefunction ansatz. Note that this energy inherently includes the kinetic energy associated with the protonic basis function centers, as derived in the Supplementary Material. Specifically, this term arises from the kinetic energy operator for the quantum protons operating on the wavefunction ansatz given by Eq. \ref{eq:Basis2}. As also shown in the Supplementary Material, this energy is the conserved quantity for RT-NEO-TDDFT Ehrenfest dynamics with the c-TPB method.  However, the conserved quantity for the original TPB method is not clearly defined. Neither the system energy, which does not include the kinetic energy associated with the protonic basis function centers, nor the extended energy, which does include this kinetic energy, is rigorously conserved in the original TPB method. 

As shown in Figure \ref{fig:ohbae}, the trajectory generated with the c-TPB method exhibits the best energy conservation among the three methods studied. Specifically, the c-TPB energy fluctuates on the order of $10^{-5}$ a.u. The trajectory generated with the FPB method exhibits larger energy fluctuations because it was generated with a timestep 10 times larger than that used for the c-TPB trajectory. \nick{This larger timestep was used for the FPB method due to its significantly greater computational expense. Comparison of the c-TPB and FPB methods with the same timestep of 0.4 a.u. shows qualitatively similar energy conservation (Figure S3). Note that the energy fluctuations tend to increase as the protonic basis function center moves more.} For the original TPB method, neither the system energy nor the extended energy were conserved, as expected. These findings are consistent with previously published results.\cite{zhao_excited_2021,10.1063/5.0255984} In terms of computational cost, Table \ref{tab:cpu} shows that the original TPB and c-TPB methods are comparable, while the FPB method is significantly more expensive because it requires many more basis functions. 

These results for the oHBA system demonstrate the accuracy, energy conservation, and computational efficiency of the c-TPB method. In comparison to the previously developed sc-TPB method, \cite{10.1063/5.0230570} which does not include the ${\bf A}$ matrix or the Lagrangian constraints, the c-TPB method is more robust because the protonic basis function centers remain constrained to the expectation values of the proton position operators. These constraints prevent drifting of the protonic basis function centers, which may lead to numerical instabilities at longer times. Such a deviation between the protonic basis function center and the proton position expectation value is shown for a trajectory of photoinduced proton transfer in oHBA generated with the sc-TPB method in the Supplementary Material (Figure S1A). Similar deviations are observed for trajectories generated with the original TPB method (Figure S1B).

\begin{table}
\caption{Average CPU time per step for the RT-NEO-TDDFT Ehrenfest dynamics simulation of the oHBA molecule with each method. The simulations were performed on 32 processors on one 2.8 GHz Intel Cascade Lake node.}
\label{tab:cpu}
\begin{ruledtabular}
\begin{tabular}{lc}
Method & Average CPU Time per step (s) \\
\hline
FPB & 873 \\
c-TPB & 291 \\
Original TPB & 307 \\
\end{tabular}
\end{ruledtabular}
\end{table}

\begin{figure}
    \centering
    \includegraphics[width=\linewidth]{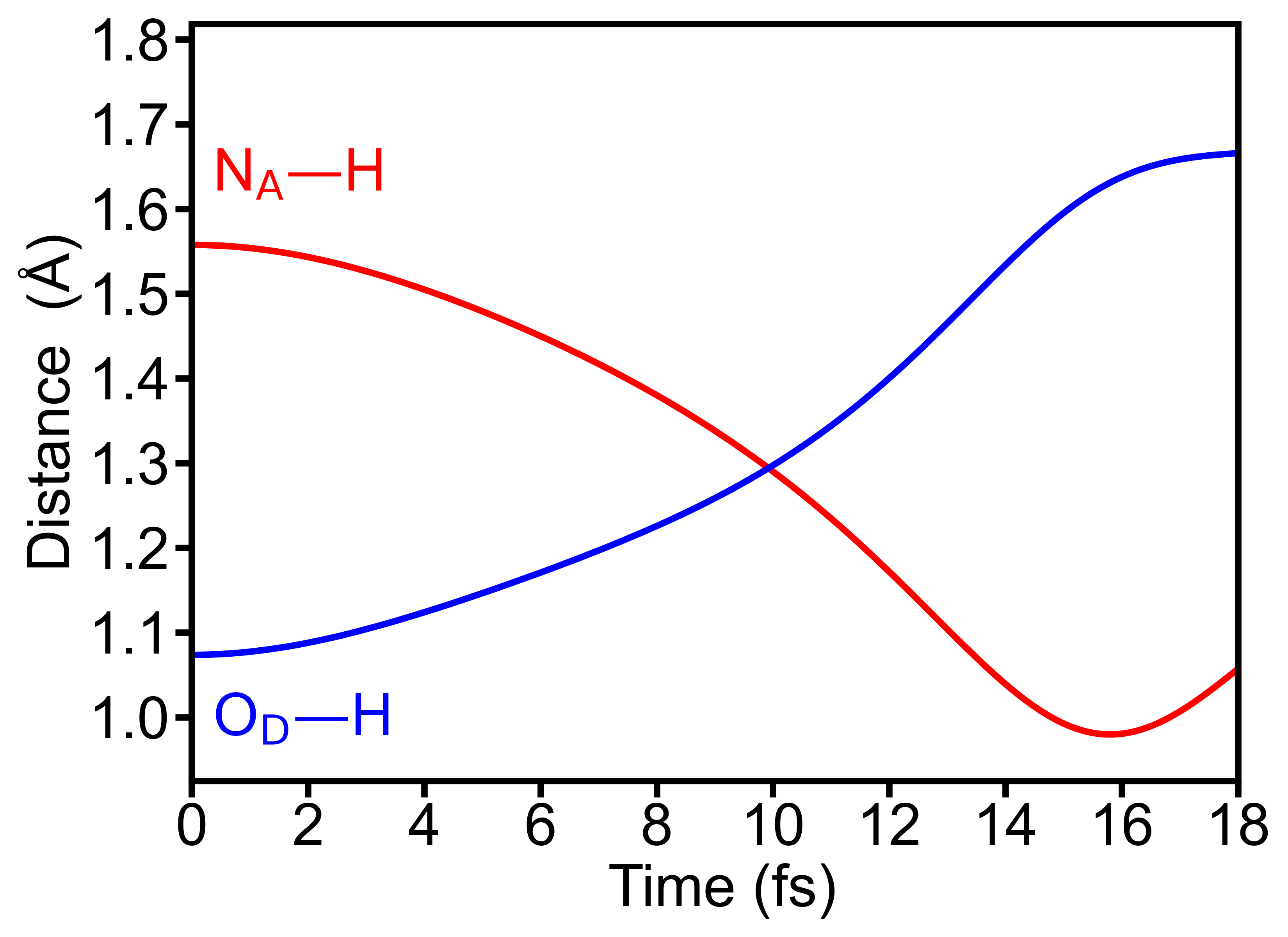}
    \caption{Distance between the transferring proton position expectation value and the donor oxygen or acceptor nitrogen atom along the trajectory for excited-state intramolecular double proton transfer in BP(OH)$_2$, as obtained from RT-NEO-TDDFT Ehrenfest dynamics with the c-TPB method. Both transferring protons are treated quantum mechanically, and due to symmetry the relative proton transfer distances are identical along the trajectory.}
    \label{fig:bp}
\end{figure}

\begin{figure}
    \centering
    \includegraphics[width=\linewidth]{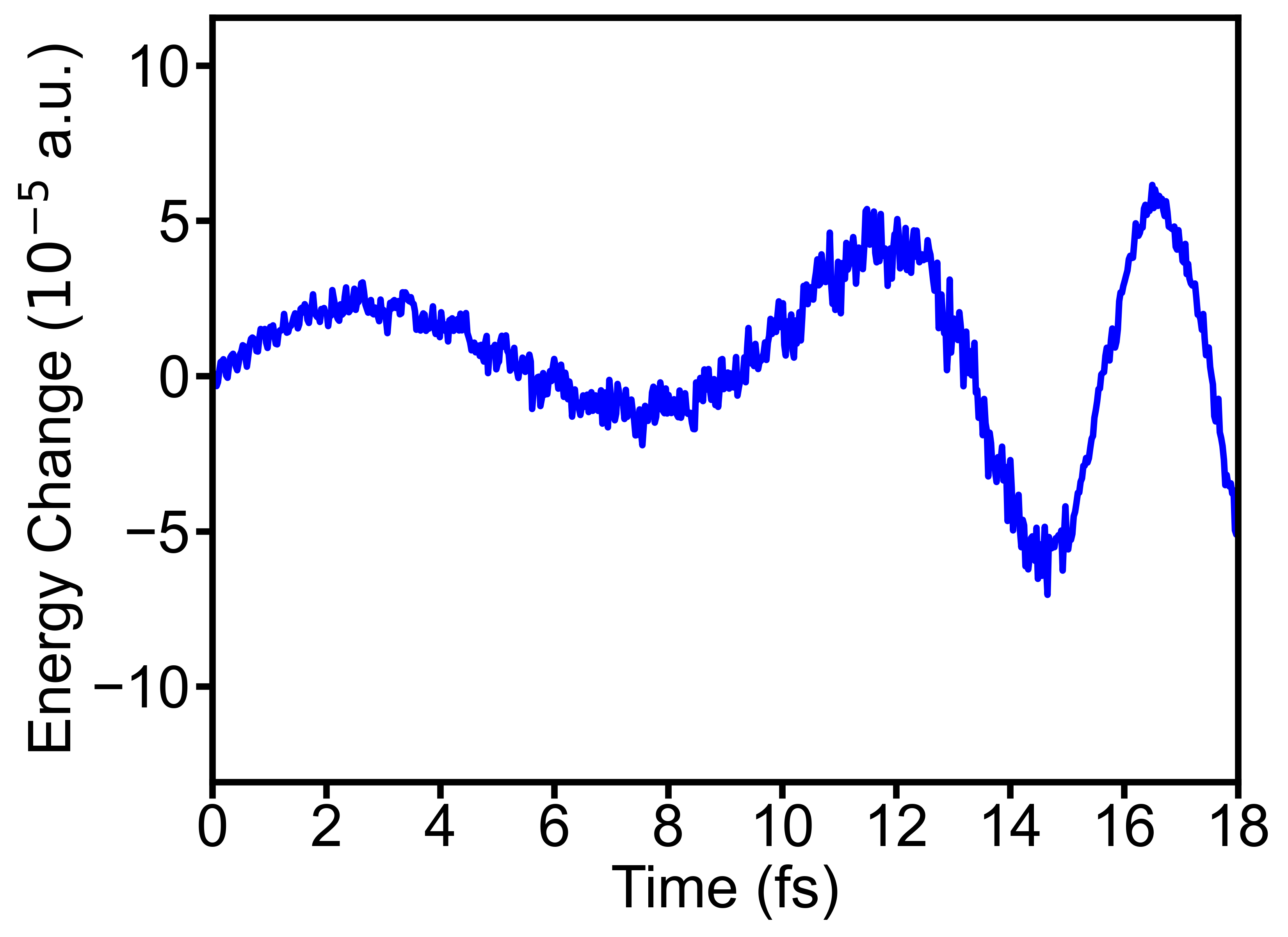}
    \caption{Energy change along the trajectory for excited-state intramolecular double proton transfer in BP(OH)$_2$, as obtained from RT-NEO-TDDFT Ehrenfest dynamics with the c-TPB method.}
    \label{fig:bpe}
\end{figure}

We also studied the excited-state intramolecular double proton transfer in the BP(OH)$_2$ molecule. Figure \ref{fig:bp} shows the distance between the transferring proton and the donor oxygen or acceptor nitrogen atom. Since BP(OH)$_2$ is a symmetric molecule, these distances are identical for both protons. The simultaneous double proton transfer occurs at \nick{around 10} fs. Figure \ref{fig:bpe} shows the energy change as a function of time along this trajectory. The energy conservation fluctuates numerically on the order of $10^{-5}$ a.u., similar to the oHBA molecule. This simulation demonstrates the ability of the c-TPB method to simulate two or more  protons transferring. Note that the smaller basis sets and larger time step used for this trajectory may limit the quantitative accuracy of the proton dynamics. However, these results provide a proof of principle for multi-proton nuclear--electronic quantum dynamics.

\section{Conclusion}

Herein, we have presented a constrained traveling proton basis (c-TPB) method that can be used with the RT-NEO-TDDFT Ehrenfest dynamics approach to accurately simulate nuclear--electronic quantum dynamics. This approach is based on equations derived from an extended Lagrangian combined with constraints ensuring that the proton position expectation value and corresponding protonic basis function center remain the same during the dynamics. This approach conserves energy, is computationally efficient, and can be used for systems with multiple quantum protons. Our applications of this approach to molecular systems highlight the use of this approach for computing vibrational spectra as well as excited-state intramolecular proton transfer dynamics for single and double proton transfer reactions.


The c-TPB strategy for moving protonic basis function centers could also be combined with other RT-NEO methods. For example, a modified version of the c-TPB method could be combined with the time-dependent configuration interaction method to simulate hydrogen tunneling dynamics and vibronic spectra involving double electron--proton excitations.\cite{10.1063/5.0243394,doi:10.1021/acs.jpclett.4c00805} Hydrogen tunneling dynamics could also be simulated by combining this strategy with the multistate DFT method.\cite{doi:10.1021/acs.jctc.2c00938,doi:10.1021/acs.jctc.4c00737} These RT-NEO dynamics methods will enable simulations of a wide range of chemical and biological processes, including proton-coupled electron transfer and proton transport via the Grotthuss mechanism.


\section*{SUPPLEMENTARY MATERIAL}

The supplementary material contains derivations of the equations for the original TPB and c-TPB methods, analysis of a harmonic oscillator model, additional analyses of oHBA trajectories and vibrational frequencies, geometries, and an example Q-Chem input file.

\begin{acknowledgments}
This work was supported by the National Science Foundation Grant No. CHE-2408934. This material is based upon work supported
by the National Science Foundation Graduate Research Fellowship under Grant No. DGE-2444107. Any opinions, findings, and conclusions or recommendations expressed in this material are those of the author(s) and do not necessarily reflect the views of the National Science Foundation. We would like to thank Dr. Mathew Chow, Dr. Chiara Donatella Aieta, Dr. Tao E. Li, Dr. Jonathan Fetherolf, Joseph Dickinson, Scott Garner, Zain Zaidi, and Millan Welman for helpful discussions.
\end{acknowledgments}

\section*{AUTHOR DECLARATIONS}

The authors have no conflicts to disclose.

\section*{Data Availability}

The data that support the findings of this study are available within this article and its Supporting Information. Also, all Q-Chem output files and raw data used to
support these findings will be openly available on Github.

\section*{References}

\bibliography{cite}

\end{document}



\title[]{\begin{center}\LARGE
Supplementary Material
\end{center}\\ \LARGE Real-time nuclear--electronic orbital time-dependent density functional theory with a constrained traveling proton basis}
\author{Nicholas J. Boyer}

\author{Sharon Hammes-Schiffer*}%

\affiliation{ 
Department of Chemistry, Princeton University, Princeton, New Jersey 08544, USA\\  \\ \small \textrm{*shs566@princeton.edu}}

\renewcommand{\thesection}{S\arabic{section}}
\renewcommand{\thefigure}{S\arabic{figure}}
\renewcommand{\thetable}{S\arabic{table}}
\renewcommand{\theequation}{S\arabic{equation}}
\renewcommand{\thepage}{S\arabic{page}}

\setcounter{section}{0}
\setcounter{figure}{0}
\setcounter{table}{0}
\setcounter{equation}{0}
\setcounter{page}{1}

\maketitle

\tableofcontents


\newpage
\section{\MakeUppercase{Definitions of quantities in main text}}
\label{sec:eqns}


The effective potentials are given by~\cite{PhysRevLett.101.153001}
\begin{equation}
U_{\mathrm{\mathrm{eff}}}^{\mathrm{p}}(\mathbf{r}^{\mathrm{p}}_1, t) = \textstyle\sum_I^{N_{\mathrm{c}}}\frac{Z_I}{| \mathbf{r}^{\mathrm{p}}_1- \mathbf R_I(t)| }+\int  \frac{\rho^{\mathrm{p}}(\mathbf{r}^{\mathrm{p}}_2, t)}{|\mathbf{r}^{\mathrm{p}}_1-\mathbf{r}^{\mathrm{p}}_2|} d \mathbf{r}^{\mathrm{p}}_2 - \int  \frac{\rho^{\mathrm{e}} (\mathbf{r}^{\mathrm{e}}_1, t)}{|\mathbf{r}^{\mathrm{p}}_1-\mathbf{r}^{\mathrm{e}}_1|} d \mathbf{r}^{\mathrm{e}}_1 + \frac{\delta E_{\mathrm{epc}}[\rho^{\mathrm{e}}, \rho^{\mathrm{p}}]}{\delta \rho^{\mathrm{p}}(\mathbf{r}^{\mathrm{p}}_1, t)} + \frac{\delta E_{\mathrm{pxc}}[\rho^{\mathrm{p}}]}{\delta \rho^{\mathrm{p}}(\mathbf{r}^{\mathrm{p}}_1, t)}
\end{equation}
\begin{equation}
U_{\mathrm{\mathrm{eff}}}^{\mathrm{e}}(\mathbf{r}_1^{\mathrm{e}}, t) = -\textstyle\sum_I^{N_{\mathrm{c}}}\frac{Z_I}{| \mathbf {r}^{\mathrm{e}}_1- \mathbf R_I(t)| }+\int  \frac{\rho^{\mathrm{e}}(\mathbf{r}^{\mathrm{e}}_2, t)}{|\mathbf{r}^{\mathrm{e}}_1-\mathbf{r}^{\mathrm{e}}_2|} d \mathbf{r}^{\mathrm{e}}_2 - \int  \frac{\rho^{\mathrm{p}}(\mathbf{r}^{\mathrm{p}}_1, t)}{|\mathbf{r}^{\mathrm{e}}_1-\mathbf{r}^{\mathrm{p}}_1|} d \mathbf{r}^{\mathrm{p}}_1 + \frac{\delta E_{\mathrm{epc}}[\rho^{\mathrm{e}}, \rho^{\mathrm{p}}]}{\delta \rho^{\mathrm{e}}(\mathbf{r}^{\mathrm{e}}_1, t)} + \frac{\delta E_{\mathrm{exc}}[\rho^{\mathrm{e}}]}{\delta \rho^{\mathrm{e}}(\mathbf{r}_{1}^{\mathrm{e}})}
\end{equation}
where $\rho^{\mathrm{e}}(\mathbf r^{\mathrm{e}},t)$ is the electron density, $\rho^{\mathrm{p}}(\mathbf r^{\mathrm{p}},t)$ is the proton density, $E_{\mathrm{exc}}[\rho^{\mathrm{e}}] $ is the electron exchange--correlation functional, $E_{\mathrm{pxc}}[\rho^{\mathrm{e}}] $ is the proton exchange--correlation functional, and  $E_{\mathrm{epc}}[\rho^{\mathrm{e}},\rho^{\mathrm{p}}]$ is the electron--proton correlation functional. Moreover, $N^{\mathrm c}$ is the number of classical nuclei, and $\mathbf{R}_I$, $m_I$, and $Z_I$ are the position, mass, and charge, respectively, of the $I$th classical nucleus.


The Kohn-Sham matrices are given by~\cite{10.1063/5.0255984}
\begin{equation}
\begin{aligned}
\mathbf F^{\text{e}}(t) &= \mathbf H_{\text{core}}^{\text{e}} 
+ \mathbf J^{\text{ee}}(\mathbf P^{\text{e}}(t)) 
+ \mathbf V_{\text{xc}}^{\text{e}}(\mathbf P^{\text{e}}(t)) \\
&\quad - \mathbf J^{\text{ep}}(\mathbf P^{\text{p}}(t)) 
+ \mathbf V_{\text{c}}^{\text{ep}}(\mathbf P^{\text{e}}(t),\mathbf P^{\text{p}}(t))
\end{aligned}
\end{equation}
\begin{equation} 
\begin{aligned} \mathbf F^{\text{p}}(t) &  = \mathbf H_{\text{core}}^{\text{p}} + \mathbf J^{\text{pp}}(\mathbf P^{\text{p}}(t)) + \mathbf V_{\text{xc}}^{\text{p}}(\mathbf P^{\text{p}}(t)) \\  & \hspace{1em}- \mathbf J^{\text{pe}}(\mathbf P^{\text{e}}(t)) + \mathbf V_{\text{c}}^{\text{pe}}(\mathbf P^{\text{p}}(t),\mathbf P^{\text{e}}(t))\hspace{0.5em}. \\  \end{aligned}
\end{equation}
Here, $\mathbf H_{\text{core}}^{\text{e/p}}$ is the electronic/protonic core Hamiltonian, $\mathbf J^{\text{ee/pp}}(\mathbf P^{\text{e/p}}(t))$ is the electronic/protonic Coulomb interaction term, $\mathbf V_{\text{xc}}^{\text{e/p}}(\mathbf P^{\text{e/p}}(t))$ is the electronic/protonic exchange--correlation term, $\mathbf J^{\text{ep/pe}}(\mathbf P^{\text{p/e}}(t))$ is the electron--proton Coulomb interaction term, and $\mathbf V_{\text{c}}^{\text{ep/pe}}(\mathbf P^{\text{e/p}}(t),\mathbf P^{\text{p/e}}(t))$ is the electron--proton correlation term. These terms depend on $\mathbf P^{\text{e}}$ and $\mathbf P^{\text{p}}$, the electronic and protonic density matrices.


\section{DERIVATION OF ORIGINAL TPB EQUATIONS}

Starting with Eq. 9 in the manuscript,
\begin{equation}
\begin{aligned}
\langle \phi_j^{\mathrm{p}}(\mathbf{r}^{\mathrm{p}} - \mathbf{R}^{\mathrm{p}}(t)) 
\mid i \frac{\partial}{\partial t} \textstyle\sum_k c_k^{\mathrm{p}}(t) 
\mid \phi_k^{\mathrm{p}}(\mathbf{r}^{\mathrm{p}} - \mathbf{R}^{\mathrm{p}}(t)) \rangle \\
= \langle \phi_j^{\mathrm{p}}(\mathbf{r}^{\mathrm{p}} - \mathbf{R}^{\mathrm{p}}(t)) 
\mid \hat{F}^{\mathrm{p}}(\mathbf{r}^{\mathrm{p}}, t) \textstyle\sum_k c_k^{\mathrm{p}}(t) 
\mid \phi_k^{\mathrm{p}}(\mathbf{r}^{\mathrm{p}} - \mathbf{R}^{\mathrm{p}}(t)) \rangle\hspace{0.5em}.
\end{aligned}
\end{equation}
\noindent
Next we calculate the derivatives on the left side of this equation:
\begin{equation} \begin{aligned}
 &\langle\phi_j^{\mathrm{p}}(\mathbf{r^{\mathrm{p}}} - \mathbf{R}^{\mathrm{p}}(t))|i  \textstyle\sum_k \left(\frac{\partial }{\partial t}c_k^{\mathrm{p}}(t) \right)|\phi_k^{\mathrm{p}}(\mathbf{r^{\mathrm{p}}} - \mathbf{R}^{\mathrm{p}}(t))\rangle\\ &+ \langle\phi_j^{\mathrm{p}}(\mathbf{r^{\mathrm{p}}} - \mathbf{R}^{\mathrm{p}}(t))|i  \textstyle\sum_k c_k^{\mathrm{p}}(t) \frac{\partial }{\partial \mathbf{R}^{\mathrm{p}}(t) }|\phi_k^{\mathrm{p}}(\mathbf{r^{\mathrm{p}}} - \mathbf{R}^{\mathrm{p}}(t))\rangle \dot{ \mathbf{R}}^{\mathrm{p}}(t) \\ &= \langle\phi_j^{\mathrm{p}}(\mathbf{r^{\mathrm{p}}} - \mathbf{R}^{\mathrm{p}}(t))|\hat{F}^{\mathrm{p}}(\mathbf{r^{\mathrm{p}}}, t) \textstyle\sum_k c_k^{\mathrm{p}}(t) |\phi_k^{\mathrm{p}}(\mathbf{r^{\mathrm{p}}} - \mathbf{R}^{\mathrm{p}}(t))\rangle\hspace{0.5em}.
\end{aligned} \end{equation}

In this work, we assume that the electronic and protonic basis sets are orthogonalized, and therefore the overlap matrix is the identity matrix.  Using the definitions of the overlap, Kohn-Sham, coefficient, and $\boldsymbol{\tau}$ matrices leads to the following relations:
\begin{equation}
\langle\phi_j^{\mathrm{p}}(\mathbf{r^{\mathrm{p}}} - \mathbf{R}^{\mathrm{p}}(t))|i  \textstyle\sum_k \left( \frac{\partial }{\partial t}c_k^{\mathrm{p}}(t) \right)|\phi_k^{\mathrm{p}}(\mathbf{r^{\mathrm{p}}} - \mathbf{R}^{\mathrm{p}}(t))\rangle = \textstyle\sum_ki\mathbf{S}_{jk}\dot c_k^{\mathrm{p}}(t) \rightarrow i \frac{\partial}{\partial t}\mathbf{C}^{\mathrm{p}} (t)
\end{equation}
\begin{equation}
\langle\phi_j^{\mathrm{p}}(\mathbf{r^{\mathrm{p}}} - \mathbf{R}^{\mathrm{p}}(t))|i  \textstyle\sum_k c_k^{\mathrm{p}}(t) \frac{\partial }{\partial \mathbf{R}^{\mathrm{p}}(t) }|\phi_k^{\mathrm{p}}(\mathbf{r^{\mathrm{p}}} - \mathbf{R}^{\mathrm{p}}(t))\rangle \dot{ \mathbf{R}}^{\mathrm{p}}(t) = \textstyle\sum_ki\mathbf{\boldsymbol{\tau}}_{jk}(t)c_k^{\mathrm{p}}(t) \rightarrow i \mathbf{\boldsymbol{\tau}} (t) \mathbf{C}^{\mathrm{p}} (t)
\end{equation}
\begin{equation}
\langle\phi_j^{\mathrm{p}}(\mathbf{r^{\mathrm{p}}} - \mathbf{R}^{\mathrm{p}}(t))|\hat{F}^{\mathrm{p}}(\mathbf{r^{\mathrm{p}}}, t) \textstyle\sum_k c_k^{\mathrm{p}}(t) |\phi_k^{\mathrm{p}}(\mathbf{r^{\mathrm{p}}} - \mathbf{R}^{\mathrm{p}}(t))\rangle = \textstyle\sum_k\mathbf{F}^{\mathrm{p}}_{jk}( t)c_k^{\mathrm{p}}(t) \rightarrow \mathbf{F}^{\mathrm{p}} (t) \mathbf{C}^{\mathrm{p}} (t)
\end{equation}
\begin{equation}
i \frac{\partial}{\partial t}\mathbf{C}^{\mathrm{p}} (t) +i \mathbf{\boldsymbol{\tau}} (t) \mathbf{C}^{\mathrm{p}} (t)=  \mathbf{F}^{\mathrm{p}}( t)\mathbf{C}^{\mathrm{p}} (t)\hspace{0.5em}.
\label{eq:11}
\end{equation}

Eq. \ref{eq:11} is Eq. 10 in the main text. Since the protonic density matrix is $\mathbf{P}^{\mathrm{p}}(t) = \mathbf{C}^{\mathrm{p}} (t) {\mathbf{C}^{\mathrm{p}}}^ {\dagger} (t)$, this equation can be rearranged to be
\begin{equation}
i \frac{\partial }{\partial t} \mathbf{P}^{\mathrm{p}}(t) = [\mathbf{F}^{\mathrm{p}}(t), \mathbf{P}^{\mathrm{p}}(t)] - i\left( \mathbf{\boldsymbol{\tau}} (t) \mathbf{P}^{\mathrm{p}}(t) +\mathbf{P}^{\mathrm{p}}(t)\mathbf{\boldsymbol{\tau}}^{\dagger}(t) \right)
\label{eq:otpb}
\end{equation}

\noindent
which is Eq. 12 in the manuscript.

\section{DERIVATION OF BASIC CONSTRAINED TPB EQUATIONS}

Starting with Eq. 14 in the manuscript,
\begin{equation} \begin{aligned}
\langle\phi_j^{\mathrm{p}}(\mathbf{r^{\mathrm{p}}} - \mathbf{R}^{\mathrm{p}}(t))|e^{-i m_{\mathrm{p}} \dot{\mathbf{R}}^{\mathrm{p}}(t) \cdot \mathbf{r^{\mathrm{p}}}} i \frac{\partial }{\partial t} \textstyle\sum_k c_k^{\mathrm{p}}(t) e^{i m_{\mathrm{p}} \dot{\mathbf{R}}^{\mathrm{p}}(t) \cdot \mathbf{r^{\mathrm{p}}}}  |\phi_k^{\mathrm{p}}(\mathbf{r^{\mathrm{p}}} - \mathbf{R}^{\mathrm{p}}(t))\rangle \\= \langle\phi_j^{\mathrm{p}}(\mathbf{r^{\mathrm{p}}} - \mathbf{R}^{\mathrm{p}}(t))|  e^{-i m_{\mathrm{p}} \dot{\mathbf{R}}^{\mathrm{p}}(t) \cdot \mathbf{r^{\mathrm{p}}}} \hat{F}^{\mathrm{p}}(\mathbf{r^{\mathrm{p}}}, t) \textstyle\sum_k c_k^{\mathrm{p}}(t) e^{i m_{\mathrm{p}} \dot{\mathbf{R}}^{\mathrm{p}}(t) \cdot \mathbf{r^{\mathrm{p}}}}|\phi_k^{\mathrm{p}}(\mathbf{r^{\mathrm{p}}} - \mathbf{R}^{\mathrm{p}}(t))\rangle\hspace{0.5em}.
\label{eq:plug}
\end{aligned} \end{equation}
\noindent
Next we compute the derivatives, including $\hat{F}^{\mathrm{p}}(\mathbf{r^{\mathrm{p}}}, t) = \frac{-1}{2 m_{\mathrm{p}}}
\nabla^2_{\mathbf{r^{\mathrm{p}}}} +U_{\mathrm{eff}}^{\mathrm{p}}(\mathbf{r^{\mathrm{p}}}, t)$:

\begin{equation} \begin{aligned}
\left\langle\phi_j^{\mathrm{p}}(\mathbf{r^{\mathrm{p}}} - \mathbf{R}^{\mathrm{p}}(t))\left| \textstyle\sum_k i \left( \frac{\partial }{\partial t}c_k^{\mathrm{p}}(t)\right)  \right|\phi_k^{\mathrm{p}}(\mathbf{r^{\mathrm{p}}} - \mathbf{R}^{\mathrm{p}}(t)) \right\rangle \\ -\left\langle\phi_j^{\mathrm{p}}(\mathbf{r^{\mathrm{p}}} - \mathbf{R}^{\mathrm{p}}(t))\left|  \textstyle\sum_k c_k^{\mathrm{p}}(t) m_{\mathrm{p}} \ddot{\mathbf{R}}^{\mathrm{p}}(t) \cdot \mathbf{r^{\mathrm{p}}}  \right|\phi_k^{\mathrm{p}}(\mathbf{r^{\mathrm{p}}} - \mathbf{R}^{\mathrm{p}}(t))\right\rangle\\ +\left\langle\phi_j^{\mathrm{p}}(\mathbf{r^{\mathrm{p}}} - \mathbf{R}^{\mathrm{p}}(t))\left|  \textstyle\sum_k c_k^{\mathrm{p}}(t)  i \frac{\partial }{\partial \mathbf{R}^{\mathrm{p}}(t)}\right|\phi_k^{\mathrm{p}}(\mathbf{r^{\mathrm{p}}} - \mathbf{R}^{\mathrm{p}}(t))\right\rangle \dot{\mathbf{R}}^{\mathrm{p}}(t) \\= \left\langle\phi_j^{\mathrm{p}}(\mathbf{r^{\mathrm{p}}} - \mathbf{R}^{\mathrm{p}}(t))\left|   \hat{F}^{\mathrm{p}}(\mathbf{r^{\mathrm{p}}}, t) \textstyle\sum_k c_k^{\mathrm{p}}(t) \right|\phi_k^{\mathrm{p}}(\mathbf{r^{\mathrm{p}}} - \mathbf{R}^{\mathrm{p}}(t))\right\rangle \\ + \left\langle\phi_j^{\mathrm{p}}(\mathbf{r^{\mathrm{p}}} - \mathbf{R}^{\mathrm{p}}(t))\left|    \textstyle\sum_k c_k^{\mathrm{p}}(t) i  \dot{\mathbf{R}}^{\mathrm{p}}(t)\frac{\partial }{\partial \mathbf{R}^{\mathrm{p}}(t)} \right|\phi_k^{\mathrm{p}}(\mathbf{r^{\mathrm{p}}} - \mathbf{R}^{\mathrm{p}}(t))\right\rangle \\+ \left\langle\phi_j^{\mathrm{p}}(\mathbf{r^{\mathrm{p}}} - \mathbf{R}^{\mathrm{p}}(t))\left|    \textstyle\sum_k c_k^{\mathrm{p}}(t) \frac{-(i m_{\mathrm{p}} \dot{\mathbf{R}}^{\mathrm{p}}(t))^2}{2m_{\mathrm{p}} }\right|\phi_k^{\mathrm{p}}(\mathbf{r^{\mathrm{p}}} - \mathbf{R}^{\mathrm{p}}(t))\right\rangle
\label{eq:plug}
\end{aligned} \end{equation}
\noindent
where $-\frac{\partial }{\partial \mathbf{r}^{\mathrm{p}}(t)} |\phi_k^{\mathrm{p}}(\mathbf{r^{\mathrm{p}}} - \mathbf{R}^{\mathrm{p}}(t))\rangle = \frac{\partial }{\partial \mathbf{R}^{\mathrm{p}}(t)} |\phi_k^{\mathrm{p}}(\mathbf{r^{\mathrm{p}}} - \mathbf{R}^{\mathrm{p}}(t))\rangle$.
\noindent
For further simplification, we use the following definition:
\begin{equation}
\mathbf{A}_{kj}(t) =  m_{\mathrm{p}}\langle\phi_k^{\mathrm{p}}(\mathbf{r^{\mathrm{p}}} - \mathbf{R}^{\mathrm{p}}(t))|\mathbf{r^{\mathrm{p}}} | \phi_j^{\mathrm{p}}(\mathbf{r^{\mathrm{p}}} - \mathbf{R}^{\mathrm{p}}(t))\rangle \cdot \ddot{\mathbf{R}}^{\mathrm{p}}(t)\hspace{0.5em}.
\label{eq:Amat}
\end{equation}

Using the definitions of the overlap, Kohn-Sham, coefficient, $\mathbf{A}$, and $\boldsymbol{\tau}$ matrices leads to the following relations:
\begin{equation}
-\left\langle\phi_j^{\mathrm{p}}(\mathbf{r^{\mathrm{p}}} - \mathbf{R}^{\mathrm{p}}(t))\left|  \textstyle\sum_k c_k^{\mathrm{p}}(t) m_{\mathrm{p}} \ddot{\mathbf{R}}^{\mathrm{p}}(t) \cdot \mathbf{r^{\mathrm{p}}}  \right|\phi_k^{\mathrm{p}}(\mathbf{r^{\mathrm{p}}} - \mathbf{R}^{\mathrm{p}}(t))\right\rangle = \textstyle\sum_k-\mathbf{A}^{\mathrm{p}}_{jk}( t)c_k^{\mathrm{p}}(t) \rightarrow -\mathbf{A}^{\mathrm{p}} (t) \mathbf{C}^{\mathrm{p}} (t)
\end{equation}
\begin{equation}
\begin{aligned}
&\left\langle\phi_j^{\mathrm{p}}(\mathbf{r^{\mathrm{p}}} - \mathbf{R}^{\mathrm{p}}(t))\left|    \textstyle\sum_k c_k^{\mathrm{p}}(t) \frac{-(i m_{\mathrm{p}} \dot{\mathbf{R}}^{\mathrm{p}}(t))^2}{2m_{\mathrm{p}} }\right|\phi_k^{\mathrm{p}}(\mathbf{r^{\mathrm{p}}} - \mathbf{R}^{\mathrm{p}}(t))\right\rangle = \textstyle\sum_k\mathbf{S}^{\mathrm{p}}_{jk}( t)c_k^{\mathrm{p}}(t) \frac{( m_{\mathrm{p}} \dot{\mathbf{R}}^{\mathrm{p}}(t))^2}{2m_{\mathrm{p}} } \\ & \rightarrow \frac{1}{2} m_{\mathrm{p}} (\dot{\mathbf{R}}^{\mathrm{p}}(t))^2 \mathbf{C}^{\mathrm{p}} (t)
\end{aligned}
\end{equation}
\begin{equation} \begin{aligned}
i \frac{\partial}{\partial t}\mathbf{C}^{\mathrm{p}} (t) +i \mathbf{\boldsymbol{\tau}} (t) \mathbf{C}^{\mathrm{p}} (t) - \mathbf{A}(t)\mathbf{C}^{\mathrm{p}} (t) =  \mathbf{F}^{\mathrm{p}}( t)\mathbf{C}^{\mathrm{p}} (t)+\frac{1}{2} m_{\mathrm{p}} (\dot{\mathbf{R}}^{\mathrm{p}}(t))^2 \mathbf{C}^{\mathrm{p}} (t)+i \mathbf{\boldsymbol{\tau}} (t) \mathbf{C}^{\mathrm{p}} (t)\hspace{0.5em}.
\label{eq:16}
\end{aligned} \end{equation}
\noindent
Eq. \ref{eq:16} is Eq. 15 in the manuscript. 

Simplifying Eq. \ref{eq:16} by removing $\frac{1}{2} m_{\mathrm{p}} (\dot{\mathbf{R}}^{\mathrm{p}}(t))^2$ because it is a spatial constant with respect to ${\bf r}^{\mathrm{p}}$ leads to
\begin{equation} \begin{aligned}
i \frac{\partial}{\partial t}\mathbf{C}^{\mathrm{p}} (t)   =  \mathbf{F}^{\mathrm{p}}( t)\mathbf{C}^{\mathrm{p}} (t)+ \mathbf{A}(t)\mathbf{C}^{\mathrm{p}} (t)\hspace{0.5em}.
\label{eq:na}
\end{aligned} \end{equation}
Since the density matrix is $\mathbf{P}^{\mathrm{p}}(t) = \mathbf{C}^{\mathrm{p}} (t) {\mathbf{C}^{\mathrm{p}}}^ {\dagger} (t)$, this equation can be rearranged to be
\begin{equation}
i \frac{\partial }{\partial t} \mathbf{P}^{\mathrm{p}}(t) = [\mathbf{F}^{\mathrm{p}}(t)+\mathbf{A}(t), \mathbf{P}^{\mathrm{p}}(t)] 
\label{eq:ntpb}
\end{equation}
\noindent
which is Eq. 17 in the manuscript. 

In the c-TPB method, a constraint is applied to ensure that the proton basis function center is the same as the expectation value of the proton position operator for each quantum proton. Application of this constraint modifies the definition of ${\bf A}(t)$ and requires the solution of a Lagrange multiplier at each time step. These equations are derived from the Lagrangian formulation in the next section.

\section{\MakeUppercase{Lagrangian Formulation for Constrained TPB Equations}}

Here we derive the von Neumann equations using the Lagrangian formulation, starting from the Lagrangian given in Ref.  ~\onlinecite{10.1063/5.0230570}. The Lagrangian for a NEO-DFT system is
\begin{equation}
\begin{aligned} &L^{\text{NEO}}(t)  = \int \text{d}\mathbf r^{\mathrm{e}}\textstyle\sum_n\psi_{n}^{\mathrm{e}*}(\mathbf r^{\mathrm{e}},t)\left[i\frac{\partial }{\partial t} + \frac{1}{2}\nabla_{\mathbf r^{\mathrm{e}}}^2\right]\psi_{n}^{\mathrm{e}}(\mathbf r^{\mathrm{e}},t)  - \frac{1}{2}\int \text{d}\mathbf r^{\mathrm{e}}\text{d}\mathbf r^{\mathrm{e}\prime}\frac{1}{| \mathbf r^{\mathrm{e}}- \mathbf r^{\mathrm{e}\prime}| }\rho^{\mathrm{e}}(\mathbf r^{\mathrm{e}},t)\rho^{\mathrm{e}}(\mathbf r^{\mathrm{e}\prime},t)- E_{\mathrm{exc}}[\rho^{\mathrm{e}}] \\ \\ & \hspace{1em} + \int \text{d}\mathbf r^{\mathrm{p}}\textstyle\sum_n\psi_n^{\mathrm{p}*}(\mathbf r^{\mathrm{p}},t)\left[i\frac{\partial }{\partial t} + \frac{1}{2m_{\mathrm{p}}}\nabla_{\mathbf r^{\mathrm{p}}}^2\right]\psi_n^{\mathrm{p}}(\mathbf r^{\mathrm{p}},t)- \frac{1}{2}\int \text{d}\mathbf r^{\mathrm{p}}\text{d}\mathbf r^{\prime \mathrm{p}}\frac{1}{| \mathbf r^{\mathrm{p}}- \mathbf r^{\prime \mathrm{p}}| }\rho^{\mathrm{p}}(\mathbf r^{\mathrm{p}},t)\rho^{\mathrm{p}}(\mathbf r^{\prime \mathrm{p}},t) - E_{\mathrm{pxc}}[\rho^{\mathrm{p}}]
 \\ \\ & \hspace{1em} + \int \text{d}\mathbf r^{\mathrm{e}}\text{d}\mathbf r^{\mathrm{p}}\frac{1}{| \mathbf r^{\mathrm{e}}- \mathbf r^{\mathrm{p}}| }\rho^{\mathrm{e}}(\mathbf r^{\mathrm{e}},t)\rho^{\mathrm{p}}(\mathbf r^{\mathrm{p}},t)- E_{\mathrm{epc}}[\rho^{\mathrm{e}},\rho^{\mathrm{p}}] + \textstyle\sum_I^{N_{\mathrm{c}}} \frac{1}{2}M_I \dot{\mathbf R}_I^2(t)- \textstyle\sum_{I < J}^{N_{\mathrm{c}}}\frac{Z_IZ_J}{| \mathbf R_I(t)- \mathbf R_J(t)| } \\  \\ & \hspace{1em} + \int \text{d}\mathbf r^{\mathrm{e}}\rho^{\mathrm{e}}(\mathbf r^{\mathrm{e}},t)\textstyle\sum_I^{N_{\mathrm{c}}}\frac{Z_I}{| \mathbf r^{\mathrm{e}}- \mathbf R_I(t)| }- \int \text{d}\mathbf r^{\mathrm{p}}\rho^{\mathrm{p}}(\mathbf r^{\mathrm{p}},t)\textstyle\sum_I^{N_{\mathrm{c}}}\frac{Z_I}{| \mathbf r^{\mathrm{p}}- \mathbf R_I(t)| } \\  \end{aligned}
\end{equation}
\noindent
where $\psi_n^e(\mathbf{r}^{\mathrm{e}}, t)$ is the $n$th electron orbital and $\psi_n^p(\mathbf{r}^{\mathrm{p}}, t)$ is the $n$th proton orbital.

Using the wavefunction ansatz $\psi^{\mathrm{p}}_n(\mathbf{r^{\mathrm{p}}}, t) =  e^{i m_{\mathrm{p}} \dot{\mathbf{R}}_n^{\mathrm{p}}(t) \cdot \mathbf{r^{\mathrm{p}}}}\psi'^{\mathrm{p}}_n(\mathbf{r^{\mathrm{p}}} ,t)$,
\begin{equation}
\begin{aligned}
&\left\langle \psi_n^{\mathrm{p}}(\mathbf r^{\mathrm{p}},t) 
\left| i\frac{\partial }{\partial t} + \frac{1}{2m_{\mathrm{p}}}\nabla_{\mathbf r^{\mathrm{p}}}^2
\right| \psi_n^{\mathrm{p}}(\mathbf r^{\mathrm{p}},t) \right\rangle 
= \\&
\left\langle \psi_n^{\prime \mathrm{p}}(\mathbf r^{\mathrm{p}},t) 
\left| i\frac{\partial }{\partial t} + \frac{1}{2m_{\mathrm{p}}}\nabla_{\mathbf r^{\mathrm{p}}}^2 
- m_{\mathrm{p}} \ddot{\mathbf{R}}_n^{\mathrm{p}} \cdot \mathbf r^{\mathrm{p}} 
- m_{\mathrm{p}}\frac{(\dot{\mathbf R}_n^{\mathrm{p}})^2}{2} 
+ i \dot{\mathbf{R}}_n^{\mathrm{p}} \cdot \nabla_{\mathbf r^{\mathrm{p}}}
\right| \psi_n^{\prime \mathrm{p}}(\mathbf r^{\mathrm{p}},t)\right\rangle \hspace{0.5em}.
\end{aligned}
\end{equation}

Then $L^{\text{NEO}}(t)$ can be written as:
\begin{equation}
L^{\text{NEO}} = L'^{\text{NEO}} 
-\textstyle\sum_n \Big(
m_{\mathrm{p}}\ddot{\mathbf{R}}_n^{\mathrm{p}} \cdot 
\langle \psi_n^{\prime \mathrm{p}}| \mathbf r^{\mathrm{p}} |\psi_n^{\prime \mathrm{p}} \rangle
+ m_{\mathrm{p}}\frac{(\dot{\mathbf R}_n^{\mathrm{p}})^2}{2}
- i \dot{\mathbf{R}}_n^{\mathrm{p}} \cdot 
\langle \psi_n^{\prime \mathrm{p}}| \nabla_{\mathbf r^{\mathrm{p}}} |\psi_n^{\prime \mathrm{p}} \rangle
\Big)
\label{eq:Lnew}
\end{equation}

\noindent
where $L'^{\text{NEO}}$ is the Lagrangian for $\psi'^{\mathrm{p}}_n(\mathbf{r^{\mathrm{p}}} ,t)$. Eq. \ref{eq:Lnew} is the same as Eq. 23 in Ref. ~\onlinecite{10.1063/5.0230570}. As shown in Ref. ~\onlinecite{10.1063/5.0230570}, the $\ddot{\mathbf{R}}^{\mathrm{p}}_n$ term leads to a numerical instability. To alleviate this issue, we add $\mathbf{\boldsymbol{\lambda}}_n \cdot (\mathbf{R}_n^{\mathrm{p}} -\langle \psi'^{\mathrm{p}}_n| \mathbf r^{\mathrm{p}} |\psi'^{\mathrm{p}}_n \rangle )$, a Lagrange multiplier term that vanishes when the constraint is satisfied, leading to the following Lagrangian:
\begin{equation}
\begin{aligned}
L^{\text{NEO c-TPB}} =& L'^{\text{NEO}} 
-\textstyle\sum_n \Big(
-\mathbf{\boldsymbol{\lambda}}_n \cdot \big(\mathbf{R}_n^{\mathrm{p}} 
- \langle \psi_n^{\prime \mathrm{p}}| \mathbf r^{\mathrm{p}} |\psi_n^{\prime \mathrm{p}} \rangle \big)
+ m_{\mathrm{p}} \ddot{\mathbf{R}}_n^{\mathrm{p}}
\cdot \langle \psi_n^{\prime \mathrm{p}}| \mathbf r^{\mathrm{p}} |\psi_n^{\prime \mathrm{p}} \rangle
+ m_{\mathrm{p}} \frac{(\dot{\mathbf R}_n^{\mathrm{p}})^2}{2}
\\ &- i \dot{\mathbf{R}}_n^{\mathrm{p}} \cdot 
\langle \psi_n^{\prime \mathrm{p}}| \nabla_{\mathbf r^{\mathrm{p}}} |\psi_n^{\prime \mathrm{p}} \rangle
\Big)\hspace{0.5em}.
\end{aligned}
\end{equation}
The Lagrange multiplier $\boldsymbol{\lambda}_n$ is chosen such that the $\mathbf{R}_n^{\mathrm{p}} =\langle \psi'^{\mathrm{p}}_n| \mathbf r^{\mathrm{p}} |\psi'^{\mathrm{p}}_n \rangle $, and therefore $m_{\mathrm{p}}\ddot{\mathbf{R}}^{\mathrm{p}}_n\cdot \langle \psi'^{\mathrm{p}}_n| \mathbf r^{\mathrm{p}} |\psi'^{\mathrm{p}}_n \rangle = m_{\mathrm{p}}\ddot{\mathbf{R}}^{\mathrm{p}}_n \cdot \mathbf{R}^{\mathrm{p}}_n =\frac{d}{dt}(m_{\mathrm{p}}\dot{\mathbf{R}}^{\mathrm{p}}_n \cdot \mathbf{R}^{\mathrm{p}}_n) -m_{\mathrm{p}}(\dot{\mathbf{R}}^{\mathrm{p}}_n)^2$, leading to
\begin{equation}
\begin{aligned}
L^{\text{NEO c-TPB}} =& L'^{\text{NEO}} 
-\textstyle\sum_n \Big(
-\mathbf{\boldsymbol{\lambda}}_n \cdot \big(\mathbf{R}_n^{\mathrm{p}} 
- \langle \psi_n^{\prime \mathrm{p}}| \mathbf r^{\mathrm{p}} |\psi_n^{\prime \mathrm{p}} \rangle \big)
+ \frac{d}{dt}(m_{\mathrm{p}}\dot{\mathbf{R}}^{\mathrm{p}}_n \cdot \mathbf{R}^{\mathrm{p}}_n)
- m_{\mathrm{p}} \frac{(\dot{\mathbf R}_n^{\mathrm{p}})^2}{2}
\\ &- i \dot{\mathbf{R}}_n^{\mathrm{p}} \cdot
\langle \psi_n^{\prime \mathrm{p}}| \nabla_{\mathbf r^{\mathrm{p}}} |\psi_n^{\prime \mathrm{p}} \rangle
\Big)\hspace{0.5em}.
\label{eq:lctpb}
\end{aligned}
\end{equation}

The generalized Euler-Lagrange equation is
\begin{equation}
\frac{\partial L^{\text{NEO c-TPB}}}{\partial \mathbf{R}^{\mathrm{p}}_n} - \frac{d}{dt} \left(\frac{\partial L^{\text{NEO c-TPB}}}{\partial \dot {\mathbf{R}}^{\mathrm{p}}_n}\right) +  \frac{d^2}{dt^2} \left(\frac{\partial L^{\text{NEO c-TPB}}}{\partial \ddot {\mathbf{R}}^{\mathrm{p}}_n}\right) = 0\hspace{0.5em}.
\label{eq:gel}
\end{equation}
\noindent
Substituting the $\frac{d}{dt}(m_{\mathrm{p}}\dot{\mathbf{R}}^{\mathrm{p}}_n\cdot\mathbf{R}^{\mathrm{p}}_n) $ term in $L^{\text{NEO c-TPB}}$ from Eq. \ref{eq:lctpb} into Eq. \ref{eq:gel} yields
\begin{equation}m_{\mathrm{p}}\ddot{\mathbf{R}}^{\mathrm{p}}_n- 2 m_{\mathrm{p}}\ddot{\mathbf{R}}^{\mathrm{p}}_n+ m_{\mathrm{p}}\ddot{\mathbf{R}}^{\mathrm{p}}_n= 0\hspace{0.5em}.
\end{equation}
\noindent
This relation demonstrates that $\frac{d}{dt}(m_{\mathrm{p}}\dot{\mathbf{R}}^{\mathrm{p}}_n\cdot\mathbf{R}^{\mathrm{p}}_n)$ does not change the dynamics of the Lagrangian and therefore can be removed. Thus, the complete $L^{\text{NEO c-TPB}}$ can be expressed as
\begin{equation}
\begin{aligned}
&L^{\text{NEO c-TPB}}(t)  = \int \text{d}\mathbf r^{\mathrm{e}}\textstyle\sum_n\psi_{n}^{\mathrm{e}*}(\mathbf r^{\mathrm{e}},t)\left[i\frac{\partial }{\partial t} + \frac{1}{2}\nabla_{\mathbf r^{\mathrm{e}}}^2\right]\psi_{n}^{\mathrm{e}}(\mathbf r^{\mathrm{e}},t)  - \frac{1}{2}\int \text{d}\mathbf r^{\mathrm{e}}\text{d}\mathbf r^{\mathrm{e}\prime}\frac{1}{| \mathbf r^{\mathrm{e}}- \mathbf r^{\mathrm{e}\prime}| }\rho^{\mathrm{e}}(\mathbf r^{\mathrm{e}},t)\rho^{\mathrm{e}}(\mathbf r^{\mathrm{e}\prime},t)\\ \\ &\hspace{1em}- E_{\mathrm{exc}}^{\mathrm{e}}[\rho^{\mathrm{e}}]   + \int \text{d}\mathbf r^{\mathrm{p}}\textstyle\sum_n\psi_n'^{\mathrm{p}*}(\mathbf r^{\mathrm{p}},t)\left[i\frac{\partial }{\partial t} + \frac{1}{2m_{\mathrm{p}}}\nabla_{\mathbf r^{\mathrm{p}}}^2 
+ \mathbf{\boldsymbol{\lambda}}_n(t)\cdot\big(\mathbf{R}_n^{\mathrm{p}}(t) - \mathbf r^{\mathrm{p}}\big)
+ i \dot{\mathbf{R}}_n^{\mathrm{p}}(t)\cdot\nabla_{\mathbf r^{\mathrm{p}}}\right]\psi_n'^{\mathrm{p}}(\mathbf r^{\mathrm{p}},t)\\ \\ & \hspace{1em} - E_{\mathrm{pxc}}^{\mathrm{p}}[\rho^{\mathrm{p}}] - \frac{1}{2}\int \text{d}\mathbf r^{\mathrm{p}}\text{d}\mathbf r^{\prime \mathrm{p}}\frac{1}{| \mathbf r^{\mathrm{p}}- \mathbf r^{\prime \mathrm{p}}| }\rho^{\mathrm{p}}(\mathbf r^{\mathrm{p}},t)\rho^{\mathrm{p}}(\mathbf r^{\prime \mathrm{p}},t) 
+ \int \text{d}\mathbf r^{\mathrm{e}}\text{d}\mathbf r^{\mathrm{p}}\frac{1}{| \mathbf r^{\mathrm{e}}- \mathbf r^{\mathrm{p}}| }\rho^{\mathrm{e}}(\mathbf r^{\mathrm{e}},t)\rho^{\mathrm{p}}(\mathbf r^{\mathrm{p}},t) \\ \\ & \hspace{1em}- E_{\mathrm{epc}}[\rho^{\mathrm{e}},\rho^{\mathrm{p}}] + \textstyle\sum_I^{N_{\mathrm{c}}} \frac{1}{2}M_I \dot{\mathbf R}_I^2(t)- \textstyle\sum_{I < J}^{N_{\mathrm{c}}}\frac{Z_IZ_J}{| \mathbf R_I(t)- \mathbf R_J(t)| }  + \int \text{d}\mathbf r^{\mathrm{e}}\rho^{\mathrm{e}}(\mathbf r^{\mathrm{e}},t)\textstyle\sum_I^{N_{\mathrm{c}}}\frac{Z_I}{| \mathbf r^{\mathrm{e}}- \mathbf R_I(t)| } \\  \\ & \hspace{1em}- \int \text{d}\mathbf r^{\mathrm{p}}\rho^{\mathrm{p}}(\mathbf r^{\mathrm{p}},t)\textstyle\sum_I^{N_{\mathrm{c}}}\frac{Z_I}{| \mathbf r^{\mathrm{p}}- \mathbf R_I(t)| } + \textstyle\sum_n^{N_{\rm p}} m_{\mathrm{p}} \frac{(\dot{\mathbf R}_n^{\mathrm{p}})^2}{2}\hspace{0.5em}. \\ 
\end{aligned}
\end{equation}

Using the principle of least action, the equations of motion for all degrees of freedom can be solved from this Lagrangian, where the action is $A = \int_{t_0}^{t_1} L(t)dt$. The classical degrees of freedom can be solved via the Euler-Lagrange equation,
\begin{equation}
M_I \ddot {\mathbf{R}}_I(t) = \nabla_{\mathbf R_I } L^{\text{NEO c-TPB}}(t) = -  \nabla_{\mathbf R_I } E(\mathbf{P}^{\mathrm{e}}(t), \mathbf{P}^{\mathrm{p}}(t), \{\mathbf{R}_I(t), \mathbf{P}_I(t)\})\hspace{0.5em}.
\end{equation}
Here, the total energy $E$ is defined in Eq. 19 of the main text and also below.
Similarly, the proton basis function center equation of motion can also be solved via the Euler-Lagrange equation,
\begin{equation}
m_{\mathrm{p}} \ddot{ \mathbf{R}}_n^{\mathrm{p}}(t) = \nabla_{\mathbf{R}_n^{\mathrm{p}}} L^{\text{NEO c-TPB}}(t) =   -\nabla_{\mathbf{R}_n^{\mathrm{p}} } E(\mathbf{P}^{\mathrm{e}}(t), \mathbf{P}^{\mathrm{p}}(t), \{\mathbf{R}_I(t), \mathbf{P}_I(t)\})\hspace{0.5em}.
\label{eq:gradients}
\end{equation}

For the quantum degrees of freedom, we use the variational principle for the action, namely
$\frac{\delta A}{\delta \psi_{n}^{\mathrm{e}*}(\mathbf r^{\mathrm{e}},t)} = 0$ 
and 
$\frac{\delta A}{\delta \psi_{n}^{\prime \mathrm{p}*}(\mathbf r^{\mathrm{p}},t)} = 0$, leading to
\begin{equation}
i \frac{\partial }{\partial t} \psi^{\mathrm{e}}_n(\mathbf{r^{\mathrm{e}}}, t) = \left(\frac{-1}{2 m_{\mathrm{e}}}
\nabla^2_{\mathbf{r^{\mathrm{e}}}} +U_{\mathrm{eff}}^{\mathrm{e}}(\mathbf{r^{\mathrm{e}}}, t) \right) \psi^{\mathrm{e}}_n(\mathbf{r^{\mathrm{e}}}, t)
\end{equation}
\begin{equation}
i \frac{\partial }{\partial t} \psi'^{\mathrm{p}}_n(\mathbf{r^{\mathrm{p}}}, t) = \left(\frac{-1}{2 m_{\mathrm{p}}}
\nabla^2_{\mathbf{r^{\mathrm{p}}}} +U_{\mathrm{eff}}^{\mathrm{p}}(\mathbf{r^{\mathrm{p}}}, t) -\mathbf{\boldsymbol{\lambda}}_n \cdot \left( \mathbf{R}_n^{\mathrm{p}}(t) - \mathbf r^{\mathrm{p}}\right)  -i  \dot{\mathbf{R}}_n^{\mathrm{p}}(t) \cdot \nabla_{\mathbf r^{\mathrm{p}} }\right) \psi'^{\mathrm{p}}_n(\mathbf{r^{\mathrm{p}}}, t)\hspace{0.5em}.
\end{equation} 

Representing the orbital for a given $n$ in terms of basis functions, we obtain  $\psi^{\mathrm{e}}_n(\mathbf{r^{\mathrm{e}}}, t) = \textstyle\sum_k c_{n,k}^{\mathrm{e}}(t)\phi_k^{\mathrm{e}}(\mathbf{r^{\mathrm{e}}})$, where the center dependence is ignored for the electronic basis functions, and $\psi'^{\mathrm{p}}_n (\mathbf{r^{\mathrm{p}}}, t) = \textstyle\sum_k c_{n, k}^{\mathrm{p}}(t)\phi_{n, k}^{\mathrm{p}}(\mathbf{r^{\mathrm{p}}}-\mathbf{R}^{\mathrm{p}}_n(t))$, leading to\
\begin{equation}
\left\langle \phi_j^{\mathrm{e}}(\mathbf{r^{\mathrm{e}}})\left|  i \frac{\partial }{\partial t} \left(\textstyle\sum_k c_{n,k}^{\mathrm{e}}(t) \right| \phi_k^{\mathrm{e}}(\mathbf{r^{\mathrm{e}}})\right\rangle \right) = \left\langle \phi_j^{\mathrm{e}}(\mathbf{r^{\mathrm{e}}})\left|\left(\frac{-1}{2 m_{\mathrm{e}}}
\nabla^2_{\mathbf{r^{\mathrm{e}}}} +U_{\mathrm{eff}}^{\mathrm{e}}(\mathbf{r^{\mathrm{e}}}, t) \right) \textstyle\sum_k c_{n,k}^{\mathrm{e}}(t) \right|\phi_k^{\mathrm{e}}(\mathbf{r^{\mathrm{e}}}) \right\rangle \
\end{equation}

\begin{multline}
\left\langle \phi_{n, j}^{\mathrm{p}}(\mathbf{r^{\mathrm{p}}}-\mathbf{R}^{\mathrm{p}}_n(t))\left|  i \frac{\partial }{\partial t} \left(\textstyle\sum_k c_{n,k}^{\mathrm{p}}(t) \right| \phi_{n ,k}^{\mathrm{p}}(\mathbf{r^{\mathrm{p}}}-\mathbf{R}^{\mathrm{p}}_n(t)) \right\rangle \right) =\\
\left\langle \phi_{n, j}^{\mathrm{p}}(\mathbf{r^{\mathrm{p}}}-\mathbf{R}^{\mathrm{p}}_n(t)) \left|  i \frac{\partial }{\partial t} \left(\textstyle\sum_k c_{n,k}^{\mathrm{p}}(t)\right) \right| \phi_{n ,k}^{\mathrm{p}}(\mathbf{r^{\mathrm{p}}}-\mathbf{R}^{\mathrm{p}}_n(t)) \right\rangle \\+ \left\langle \phi_{n, j}^{\mathrm{p}}(\mathbf{r^{\mathrm{p}}}-\mathbf{R}^{\mathrm{p}}_n(t))\left|    -i\textstyle\sum_k c_{n,k}^{\mathrm{p}}(t)  \dot{ \mathbf{R}}_n^{\mathrm{p}} \cdot \nabla_{\mathbf r^{\mathrm{p}} } \right| \phi_{n ,k}^{\mathrm{p}}(\mathbf{r^{\mathrm{p}}}-\mathbf{R}^{\mathrm{p}}_n(t)) \right\rangle  =\\ \left\langle \phi_{n, j}^{\mathrm{p}}(\mathbf{r^{\mathrm{p}}}-\mathbf{R}^{\mathrm{p}}_n(t))\left|\left(\frac{-1}{2 m_{\mathrm{p}}}
\nabla^2_{\mathbf{r^{\mathrm{p}}}} +U_{\mathrm{eff}}^{\mathrm{p}}(\mathbf{r^{\mathrm{p}}}, t) -\mathbf{\boldsymbol{\lambda}}_n \cdot (\mathbf{R}_n^{\mathrm{p}} - \mathbf r^{\mathrm{p}})  -i \dot{ \mathbf{R}}_n^{\mathrm{p}} \cdot \nabla_{\mathbf r^{\mathrm{p}} }\right) \textstyle\sum_k c_{n,k}^{\mathrm{p}}(t) \right|\phi_{n ,k}^{\mathrm{p}}(\mathbf{r^{\mathrm{p}}}-\mathbf{R}^{\mathrm{p}}_n(t)) \right\rangle\hspace{0.5em}.
\label{eq:protonic}
\end{multline} 

Since $\mathbf{\boldsymbol{\lambda}}_n\mathbf{R}_n^{\mathrm{p}}$ is a spatial constant with respect to the proton coordinate, it can be removed, and Eq. \ref{eq:protonic} can be simplified to
\begin{multline}
\left\langle \phi_{n, j}^{\mathrm{p}}(\mathbf{r^{\mathrm{p}}}-\mathbf{R}^{\mathrm{p}}_n(t))\left|  i \frac{\partial }{\partial t} \left(\textstyle\sum_k c_{n,k}^{\mathrm{p}}(t)\right) \right| \phi_{n ,k}^{\mathrm{p}}(\mathbf{r^{\mathrm{p}}}-\mathbf{R}^{\mathrm{p}}_n(t)) \right\rangle =\\ \left\langle \phi_{n, j}^{\mathrm{p}}(\mathbf{r^{\mathrm{p}}}-\mathbf{R}^{\mathrm{p}}_n(t))\left|\left(\frac{-1}{2 m_{\mathrm{p}}}
\nabla^2_{\mathbf{r^{\mathrm{p}}}} +U_{\mathrm{eff}}^{\mathrm{p}}(\mathbf{r^{\mathrm{p}}}, t) +\mathbf{\boldsymbol{\lambda}}_n\cdot  \mathbf r^{\mathrm{p}}  \right) \textstyle\sum_k c_{n,k}^{\mathrm{p}}(t) \right|\phi_{n ,k}^{\mathrm{p}}(\mathbf{r^{\mathrm{p}}}-\mathbf{R}^{\mathrm{p}}_n(t)) \right\rangle\hspace{0.5em}.
\label{eq:protonic2}
\end{multline} 

Changing the representation to matrices yields the following von Neumann equations:
\begin{equation}
i \frac{\partial }{\partial t} \mathbf{P}^{\mathrm{e}}(t) = [\mathbf{F}^{\mathrm{e}}(t), \mathbf{P}^{\mathrm{e}}(t)]
\label{eq:vonn}
\end{equation}
\begin{equation}
i \frac{\partial }{\partial t} \mathbf{P}_n^{\mathrm{p}}(t) = [\mathbf{F}_n^{\mathrm{p}}(t) + \mathbf{A}_n(t), \mathbf{P}_n^{\mathrm{p}}(t)]
\label{eq:hptd}
\end{equation}
\noindent
where 
\begin{equation}
(\mathbf{A}_n)_{kj}(t) = \langle\phi_{n, k}^{\mathrm{p}}(\mathbf{r^{\mathrm{p}}} - \mathbf{R}^{\mathrm{p}}_{n}(t))|\mathbf{r^{\mathrm{p}}} | \phi_{n, j}^{\mathrm{p}}(\mathbf{r^{\mathrm{p}}} - \mathbf{R}_{n}^{\mathrm{p}}(t))\rangle  \cdot \mathbf{\boldsymbol{\lambda}}_{n}(t) \hspace{0.5em}.
\label{eq:amat2}
\end{equation}
\noindent
Here, $n$ denotes a quantum proton orbital, and $j, k$ denote protonic basis functions. Eq. \ref{eq:Amat} is related to Eq. \ref{eq:amat2} by $\mathbf{\boldsymbol{\lambda}}_{n}(t)  = m_{\mathrm{p}}\ddot{\mathbf{R}}_n^{\mathrm{p}}(t)$. In the nuclear Hartree product representation, the quantum protons are treated as distinguishable particles, and each quantum proton will be described by Eq. \ref{eq:hptd}.

The Lagrangian multiplier, $\mathbf{\boldsymbol{\lambda}}_{n}(t) $, is chosen to satisfy the condition
\begin{equation}
\frac{d L^{\text{NEO c-TPB}}(t)}{d \mathbf{\boldsymbol{\lambda}}_{n}(t)} = 0 = \mathbf{R}_n^{\mathrm{p}}(t) -\langle \psi'^{\mathrm{p}}_n (t)| \mathbf r^{\mathrm{p}} |\psi'^{\mathrm{p}}_n (t) \rangle \hspace{0.5em}.
\end{equation}
\noindent
This equation can be solved iteratively by the procedure given in Ref. ~\onlinecite{10.1063/1.5143371}. In this procedure,
\begin{equation}
\mathbf{\boldsymbol{\lambda}}_{n}^{a+1}(t + \Delta t) =\mathbf{\boldsymbol{\lambda}}_{n}^{a}(t+ \Delta t)  - \left[\frac{d^2 L^{\text{NEO c-TPB}}(t+ \Delta t)}{d^2 \mathbf{\boldsymbol{\lambda}}_{n}^{a}(t+ \Delta t)}\right]^{-1}\frac{d L^{\text{NEO c-TPB}}(t+ \Delta t)}{d \mathbf{\boldsymbol{\lambda}}_{n}^{a}(t+ \Delta t)}
\end{equation}
\noindent
where $a$ is the iteration number and
\begin{equation}
\frac{d^2 L^{\text{NEO c-TPB}}(t+ \Delta t)}{d^2 \mathbf{\boldsymbol{\lambda}}_{n}^{a}(t+ \Delta t)} = -\frac{d}{d \mathbf{\boldsymbol{\lambda}}_{n}^{a}(t+ \Delta t)}\langle \psi'^{\mathrm{p}}_n(t+ \Delta t)| \mathbf r^{\mathrm{p}} |\psi'^{\mathrm{p}}_n (t+ \Delta t)\rangle ^a\hspace{0.5em}.
\end{equation}

The second order Taylor expansion approximation of $\langle \mathbf{r^{\mathrm{p}}} \rangle(t+\Delta t)$ is
\begin{equation}
\langle \mathbf{r^{\mathrm{p}}} \rangle(t+\Delta t) = \langle \mathbf{r^{\mathrm{p}}} \rangle(t) + \dot{\langle \mathbf{r^{\mathrm{p}}}} \rangle(t) \Delta t + \frac{1}{2}\ddot{\langle \mathbf{r^{\mathrm{p}}}} \rangle(t) \Delta t^2 
\end{equation}
\noindent
and $m_{\mathrm{p}} \ddot{\langle \mathbf{r^{\mathrm{p}}}} \rangle(t) =-\langle \nabla _{\mathbf{r^{\mathrm{p}}}} (U_{\mathrm{eff}}^{\mathrm{p}}+\mathbf{r^{\mathrm{p}}} \cdot \mathbf{\boldsymbol{\lambda}}_n^a(t))\rangle(t) $ by the Ehrenfest theorem.\cite{Ehrenfest1927} Substituting these equations into Eq. 39 leads to
\begin{equation}
\frac{d^2 L^{\text{NEO c-TPB}}(t+ \Delta t)}{d^2 \mathbf{\boldsymbol{\lambda}}_{n}^{a}(t+ \Delta t)} = \frac{\Delta t^2}{2m_{\mathrm{p}}}\hspace{0.5em}.
\end{equation}

The full equation is
\begin{equation}
\mathbf{\boldsymbol{\lambda}}_{n}^{a+1}(t + \Delta t) =\mathbf{\boldsymbol{\lambda}}_{n}^{a}(t+ \Delta t)  -\frac{2 m_{\mathrm{p}}}{\Delta t^2} (\mathbf{R}_n^{\mathrm{p}}(t+ \Delta t) -\langle \psi'^{\mathrm{p}}_n (t+ \Delta t)| \mathbf r^{\mathrm{p}} |\psi'^{\mathrm{p}}_n (t+ \Delta t) \rangle^a )\hspace{0.5em}.
\end{equation}
Defining $\mathbf{f}_n(t) m_{\mathrm{p}} = \boldsymbol{\lambda}_n$, then
\begin{equation}
\mathbf{f}_{n}^{a+1}(t + \Delta t) =\mathbf{f}_{n}^{a}(t+ \Delta t)  -\frac{2 }{\Delta t^2} (\mathbf{R}_n^{\mathrm{p}}(t+ \Delta t) -\langle \psi'^{\mathrm{p}}_n (t+ \Delta t)| \mathbf r^{\mathrm{p}} |\psi'^{\mathrm{p}}_n (t+ \Delta t) \rangle^a )\hspace{0.5em}.
\end{equation}

\section{\MakeUppercase{Conservation of Energy}}

By Noether's theorem, the energy is conserved in this Lagrangian because $\frac{\partial L^{\text{NEO c-TPB}}(t)}{\partial t}=0$.\cite{Noether01011971} 

The total energy is
\begin{equation}
\begin{aligned} &E^{\text{NEO c-TPB}}(t)  = \int \text{d}\mathbf r^{\mathrm{e}}\textstyle\sum_n\psi_{n}^{\mathrm{e}*}(\mathbf r^{\mathrm{e}},t)\left[- \frac{1}{2}\nabla_{\mathbf r^{\mathrm{e}}}^2\right]\psi_{n}^{\mathrm{e}}(\mathbf r^{\mathrm{e}},t)  + \frac{1}{2}\int \text{d}\mathbf r^{\mathrm{e}}\text{d}\mathbf r^{\mathrm{e}\prime}\frac{1}{| \mathbf r^{\mathrm{e}}- \mathbf r^{\mathrm{e}\prime}| }\rho^{\mathrm{e}}(\mathbf r^{\mathrm{e}},t)\rho^{\mathrm{e}}(\mathbf r^{\mathrm{e}\prime},t)\\ \\ &+ E_{\mathrm{exc}}^{\mathrm{e}}[\rho^{\mathrm{e}}]   + \int \text{d}\mathbf r^{\mathrm{p}}\textstyle\sum_n\psi_n'^{\mathrm{p}*}(\mathbf r^{\mathrm{p}},t)\Big[- \frac{1}{2m_{\mathrm{p}}}\nabla_{\mathbf r^{\mathrm{p}}}^2 
\Big]\psi_n'^{\mathrm{p}}(\mathbf r^{\mathrm{p}},t) + E_{\mathrm{pxc}}^{\mathrm{p}}[\rho^{\mathrm{p}}] \\ \\ & \hspace{1em} + \frac{1}{2}\int \text{d}\mathbf r^{\mathrm{p}}\text{d}\mathbf r^{\prime \mathrm{p}}\frac{1}{| \mathbf r^{\mathrm{p}}- \mathbf r^{\prime \mathrm{p}}| }\rho^{\mathrm{p}}(\mathbf r^{\mathrm{p}},t)\rho^{\mathrm{p}}(\mathbf r^{\prime \mathrm{p}},t) 
- \int \text{d}\mathbf r^{\mathrm{e}}\text{d}\mathbf r^{\mathrm{p}}\frac{1}{| \mathbf r^{\mathrm{e}}- \mathbf r^{\mathrm{p}}| }\rho^{\mathrm{e}}(\mathbf r^{\mathrm{e}},t)\rho^{\mathrm{p}}(\mathbf r^{\mathrm{p}},t) \\ \\ & \hspace{1em}+ E_{\mathrm{epc}}[\rho^{\mathrm{e}},\rho^{\mathrm{p}}] + \textstyle\sum_I^{N_{\mathrm{c}}} \frac{1}{2}M_I \dot{\mathbf R}_I^2(t)+ \textstyle\sum_{I < J}^{N_{\mathrm{c}}}\frac{Z_IZ_J}{| \mathbf R_I(t)- \mathbf R_J(t)| }  - \int \text{d}\mathbf r^{\mathrm{e}}\rho^{\mathrm{e}}(\mathbf r^{\mathrm{e}},t)\textstyle\sum_I^{N_{\mathrm{c}}}\frac{Z_I}{| \mathbf r^{\mathrm{e}}- \mathbf R_I(t)| } \\  \\ & \hspace{1em}+ \int \text{d}\mathbf r^{\mathrm{p}}\rho^{\mathrm{p}}(\mathbf r^{\mathrm{p}},t)\textstyle\sum_I^{N_{\mathrm{c}}}\frac{Z_I}{| \mathbf r^{\mathrm{p}}- \mathbf R_I(t)| } + \textstyle\sum_n^{N_{p}} \frac{1}{2} m_{\mathrm{p}} (\dot{\mathbf R}_n^{\mathrm{p}})^2 \hspace{0.5em}.
 \end{aligned}
 \label{eq:lenergy}
\end{equation}

The energy given in Eq. \ref{eq:lenergy}  is given in matrix form in Eq. 23 of the main paper.

\section{\MakeUppercase{Harmonic Oscillator Model System}}

As a proof of principle of the c-TPB approach, we use the one-dimensional harmonic oscillator system. Let the Hamiltonian be 
\begin{equation}
H = \frac{-1}{2 m_{\mathrm{p}}}\frac{\partial^2}{\partial x^2} +\frac{1}{2} m_{\mathrm{p}} \omega^2x^2
\label{eq:rqho}
\end{equation}

\noindent where $\omega = \sqrt{\frac{k}{m_{\mathrm{p}}}}$. The second-order time derivative of the expectation value is  $\frac{d^2 \langle x \rangle(t)}{dt^2} = -\omega^2 \langle x \rangle(t)$. Using Eq. \ref{eq:ntpb} to add $m_p   \ddot{\langle x \rangle } x$, the new effective Hamiltonian is

\begin{multline}
H_{\mathrm{eff}} = \frac{-1}{2 m_{\mathrm{p}}}\frac{\partial^2}{\partial x^2} +\frac{1}{2} m_{\mathrm{p}} \omega^2x^2 -  m_{\mathrm{p}} \omega^2 \langle x \rangle(t) x \\ =\frac{-1}{2 m_{\mathrm{p}}}\frac{\partial^2}{\partial x^2} +\frac{1}{2} m_{\mathrm{p}} \omega^2 (x -   \langle x \rangle(t) )^2 - \frac{1}{2} m_{\mathrm{p}} \omega^2 (\langle x \rangle(t))^2
\label{eq:effH}
\end{multline}

\noindent where the $- \frac{1}{2} m_{\mathrm{p}} \omega^2 (\langle x \rangle(t))^2$ can be ignored since it is a spatial constant in the dynamics. Since $\langle x \rangle(t)$ is the traveling proton basis function center that shifts the basis function according to Eq. 14 in the main paper, a coordinate translation is being applied between the Hamiltonian and the basis functions. This translation can be applied to the effective Hamiltonian (Eq. \ref{eq:effH}) and basis functions (Eq. 13) to obtain
\begin{equation}
    H_{\mathrm{eff}}(x+\langle x \rangle(t)) =\frac{-1}{2 m_{\mathrm{p}}}\frac{\partial^2}{\partial x^2} +\frac{1}{2} m_{\mathrm{p}} \omega^2 x^2 
    \label{eq:qho}
\end{equation}
\begin{equation}
\psi^{\mathrm{p}}(x+\langle x \rangle(t), t) = \sum_k c_k^{\mathrm{p}}(t)  e^{i m_{\mathrm{p}} \dot{\langle x \rangle}(t) x}|\phi_k^{\mathrm{p}}(x)\rangle
\label{eq:basis3}
\end{equation}

\noindent where Eq. \ref{eq:qho} is the same as the original Eq. \ref{eq:rqho}. During dynamics $\langle\psi^{\mathrm{p}}(x+\langle x \rangle(t), t)| x | \psi^{\mathrm{p}}(x+\langle x \rangle(t), t)\rangle = 0$, but $\psi^{\mathrm{p}}(x+\langle x \rangle(t),t)$ still evolves according to the known dynamics of the quantum harmonic oscillator. In practice, the translational movement of the quantum proton has been placed into the dynamics of the classical-like expectation value of the position operator, $\langle x \rangle(t)$, while the quantum dynamics of the remaining density matrix still occur without any translational movement. It is important to note that there have been no approximations made here, and the dynamics is exactly the same as Eq. \ref{eq:rqho}. This insight provides an explanation as to why the traveling proton basis performs well: it uses both the classical-like movement of the protonic basis function center from Eq. \ref{eq:gradients} and the translation-adjusted nonequilibrium quantum dynamics from Eq. \ref{eq:ntpb}.

\newpage

\section{\MakeUppercase{Additional oHBA trajectory analysis}}

\begin{figure}[h]
    \centering
    \includegraphics[width=0.5\linewidth]{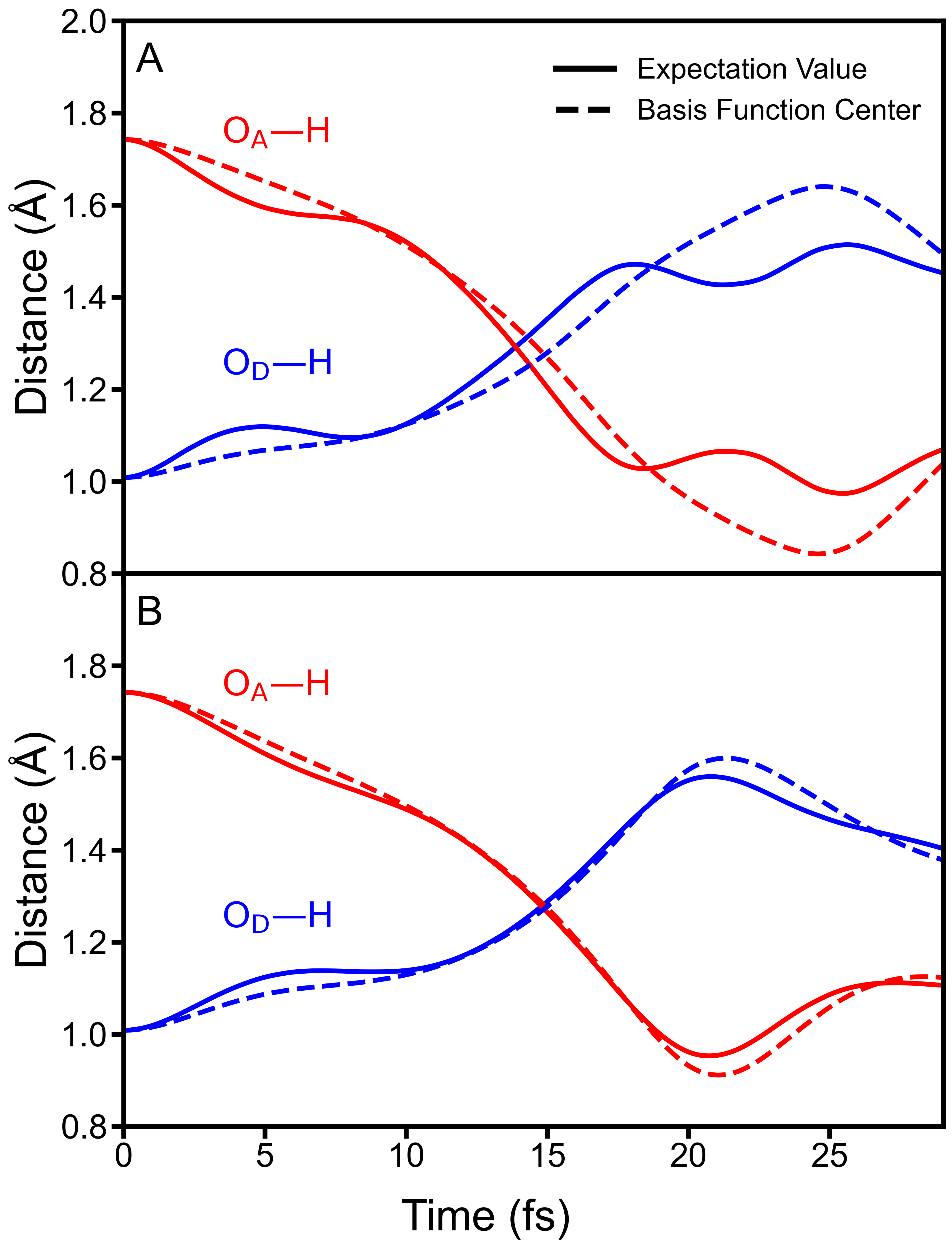}
    \caption{ Distance between the transferring proton position expectation value and the donor or acceptor oxygen atom (solid) and distance between the protonic basis function center position and the donor or acceptor oxygen atom (dashed) along the trajectories for excited-state intramolecular proton transfer in oHBA. (A) This trajectory was obtained from RT-NEO-TDDFT Ehrenfest dynamics with the previously developed sc-TPB method, where the ${\bf A}$ matrix is zero, and no Lagrangian constraints are applied. In contrast, for the c-TPB method presented herein, the ${\bf A}$ matrix is defined as in the text, and Lagrangian constraints are applied to ensure that the proton position expectation value and corresponding protonic basis function center position coincide along the entire trajectory. (B) This trajectory was obtained from RT-NEO-TDDFT Ehrenfest dynamics with the original TPB method. For the c-TPB method, the solid and dashed curves coincide exactly due to the constraint.}
    \par\noindent{Alt text:} \textit{Two-panel line plot showing O--H distances computed with proton position expectation value or protonic basis function center position versus time for excited-state proton transfer in oHBA computed using previously developed sc-TPB method and original TPB method.}
    \label{fig:sctpb}
\end{figure}
\newpage


\begin{figure}[h]
    \centering
    \includegraphics[width=0.5\linewidth]{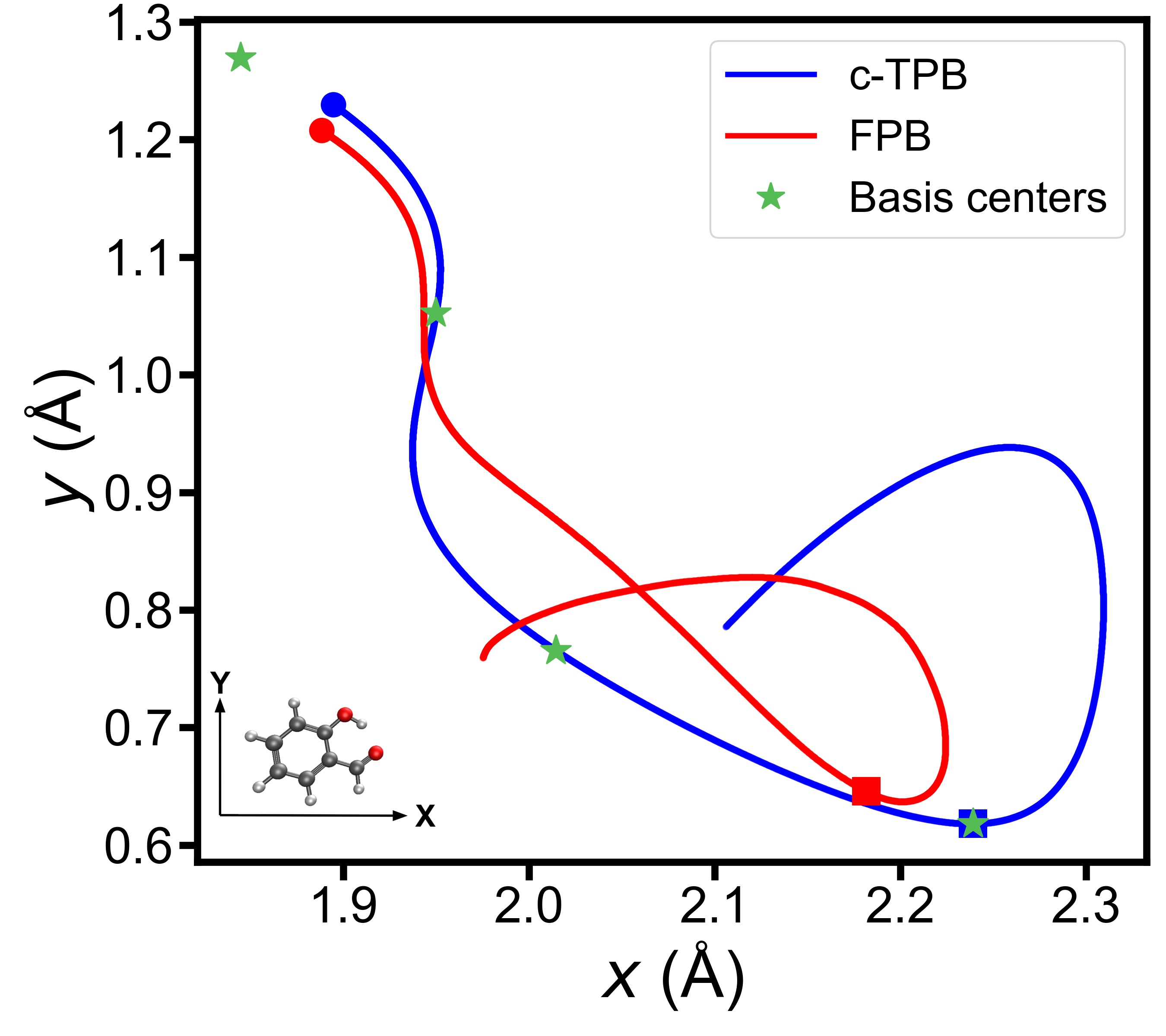}
    \caption{\nick{Plot of the $x$ and $y$ Cartesian coordinates of the proton position operator expectation value along the trajectories obtained with the c-TPB and FPB methods for excited-state intramolecular proton transfer in oHBA. The circles indicate the beginning of the trajectory. The squares indicate the location of the proton position operator expectation value at 18.5 fs. The fixed protonic basis function center positions used for the FPB method are shown with stars. The basis function center associated with the first star was positioned to ensure that the initial proton position expectation value for the FPB trajectory was similar to that for the c-TPB trajectory. This plot shows that the trajectory moves beyond the fixed protonic basis function center positions at around 18.5 fs.}}
    \par\noindent{Alt text:} \textit{Two-dimensional trajectory plot of Cartesian coordinates of transferring proton position expectation value for excited-state proton transfer in oHBA computed using c-TPB and FPB methods with fixed protonic basis function center positions marked with stars.}
    \label{fig:rev1}
\end{figure}


\begin{figure}[h]
    \centering
    \includegraphics[width=\linewidth]{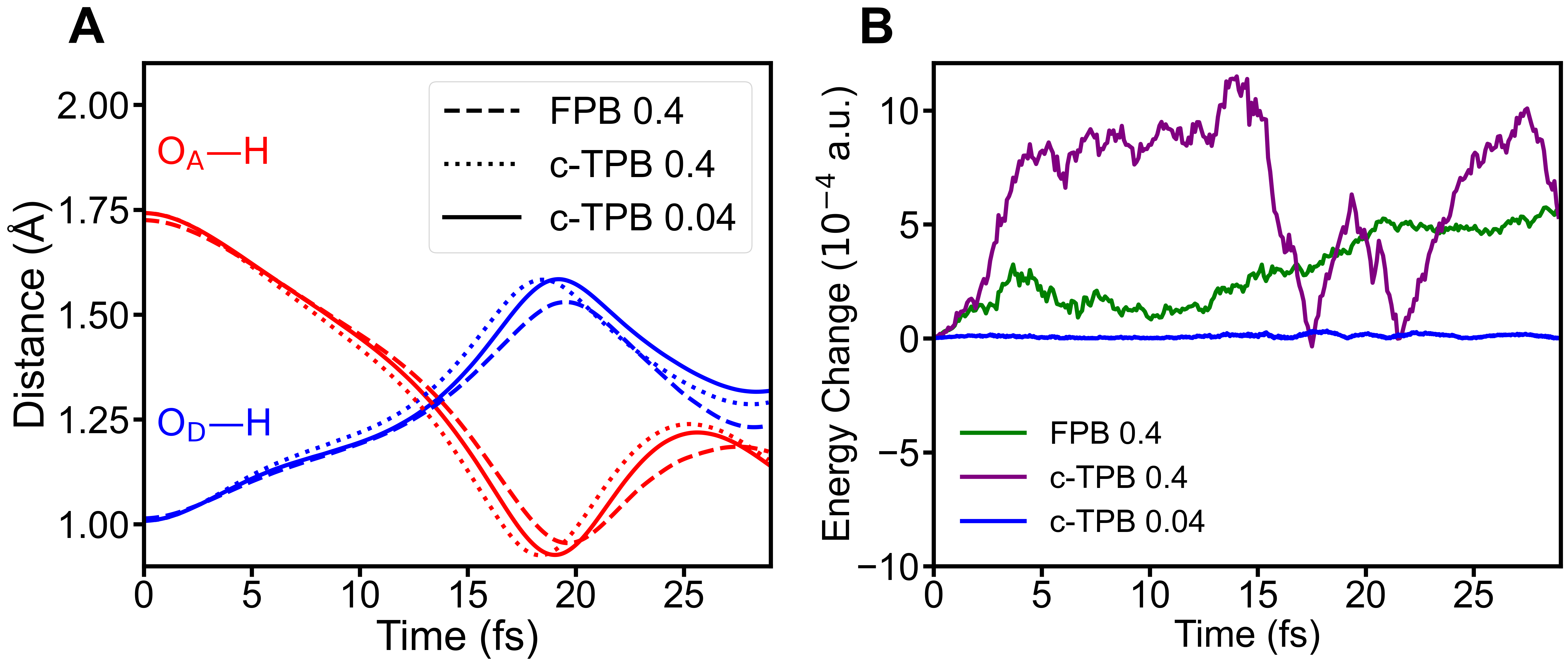}
    \caption{\nick{ (A) Distance between the transferring proton position expectation value and the donor or acceptor oxygen atom along the trajectories for excited-state intramolecular proton transfer in oHBA, as obtained from RT-NEO-TDDFT Ehrenfest dynamics with the FPB and c-TPB methods with a time step of 0.4 a.u. and the c-TPB method with a time step of 0.04 a.u. (B) Energy change along these trajectories obtained with the FPB method with a time step of 0.4 a.u. (green), the c-TPB method with a time step of 0.4 a.u. (purple), and the c-TPB method with a time step of 0.04 a.u. (blue).}}
    \par\noindent{Alt text:} \textit{Two-panel figure for excited-state proton transfer in oHBA showing O--H distance and energy change versus time computed using FPB with a 0.4~a.u.\ time step, c-TPB with a 0.4~a.u.\ time step, and c-TPB with a 0.04~a.u.\ time step.}
    \label{fig:placeholder}
\end{figure}

\newpage
\clearpage

\section{\MakeUppercase{Tables showing the convergence of vibrational frequencies}}

\begin{table*}[ht]
\centering
\footnotesize
\begin{threeparttable}

\caption{\nick{HCN, HNC, and FHF$^-$ vibrational frequencies (cm$^{-1}$) with absolute errors relative to the VPT2 reference given in parentheses. The frequencies were computed with the c-TPB method using data from the full trajectory and from the first half (h1) and second half (h2) of the trajectory to demonstrate convergence.}}

\label{tab:hcn}
\begin{tabular}{llc|ccc|ccc}
\hline
 & & & \multicolumn{3}{c|}{c-TPB} & \multicolumn{3}{c}{Original TPB} \\
\cline{4-9}
Molecule & Vibrational Mode & Ref.\ VPT2 & Full & h1 & h2 & Full & h1 & h2 \\
\hline
HCN
& CH bend    & 760  & 762 (2)    & 762 (2)    & 762 (2)    & 808 (48)    & 808 (48)     & 808 (48) \\
& CN stretch & 2176 & 2196 (20)  & 2194 (18)  & 2195 (19)  & 2242 (66)   & 2242 (66)    & 2242 (66) \\
& CH stretch & 3328 & 3345 (17)  & 3344 (16)  & 3345 (18)  & 3493 (165)  & 3493 (165)   & 3491 (164) \\
& MAE        & ---  & 13         & 12         & 13         & 93          & 93           & 93 \\
\hline
HNC
& NH bend    & 457  & 444 (13)   & 444 (13)   & 444 (13)   & 465 (8)     & 465 (8)      & 465 (8) \\
& NC stretch & 2068 & 2099 (31)  & 2100 (33)  & 2099 (31)  & 2115 (47)   & 2115 (47)    & 2115 (47) \\
& NH stretch & 3626 & 3604 (21)  & 3620 (6)   & 3613 (13)  & 3912 (287)  & 3912 (287)   & 3827 (202) \\
& MAE        & ---  & 22         & 17         & 19         & 114         & 114          & 86 \\
\hline
FHF$^-$\tnote{a}
& FF stretch & 592  & 638 (46)   & 638 (46)   & 638 (46)   & --- & --- & --- \\
& FH bend    & 1351 & 1354 (3)   & 1352 (1)   & 1352 (1)   & --- & --- & --- \\
& FH stretch & 1721 & 2006 (286) & 2006 (286) & 2006 (286) &--- & --- & --- \\
& MAE        & ---  & 111        & 111        & 111        & --- & --- & --- \\
\hline
\end{tabular}

\begin{tablenotes}
\footnotesize
\item[a] \nick{This analysis was not performed for the original TPB FHF$^-$ trajectory because numerical instabilities occurred at longer times.}
\end{tablenotes}
\par\noindent{Alt text:} \textit{Table analyzing convergence of computed vibrational frequencies for HCN, HNC, and FHF$^-$.}
\end{threeparttable}
\end{table*}

\newpage

\begin{table*}[ht]
\centering
\footnotesize
\begin{threeparttable}

\caption{\nick{H$_2$, H$_2$O, H$_2$CO, and HCOOH vibrational frequencies (cm$^{-1}$) with absolute errors relative to the VPT2 reference given in parentheses. The frequencies were computed with the c-TPB method using data from the full trajectory and from the first half (h1) and second half (h2) of the trajectory to demonstrate convergence.}}

\label{tab:multi}
\begin{tabular}{llc|ccc|ccc}
\hline
 & & & \multicolumn{3}{c|}{c-TPB} & \multicolumn{3}{c}{Original TPB} \\
\cline{4-9}
Molecule & Vibrational mode & Ref.\ VPT2 & Full & h1 & h2 & Full & h1 & h2 \\
\hline
H$_2$\tnote{a}
& HH stretch       & 4114 & 4040 (74)     & 4040 (74)  & 4040 (74)   & ---          & ---            & ---   \\
& MAE              & ---  & 74            & 74            & 74          & ---          & ---            & ---   \\
\hline
H$_2$O\tnote{a}
& HOH bend         & 1604 & 1582 (22)  & 1608 (4)      & 1585 (19)& ---          & ---            & ---   \\
& OH sym. stretch  & 3644 & 3697 (53)     & 3697 (53)     & 3697 (53)   & ---          & ---            & ---   \\
& OH asym. stretch & 3697 & 3697 (0)      & 3697 (0)      & 3697 (0)    & ---          & ---            & ---   \\
& MAE              & ---  & 25            & 19            & 24          & ---          & ---            & ---   \\
\hline
H$_2$CO\tnote{a}
& CH$_2$ wag       & 1188 & 1196 (8)      & 1195 (8)      & 1195 (8)    & ---     & 945 (243)      & ---   \\
& CH$_2$ rock         & 1247 & 1237 (11)  & 1249 (2)      & 1230 (17)   & ---   & 945 (302)      & ---  \\
& CH$_2$ scissors  & 1517 & 1483 (35)  & 1483 (35)  & 1533 (16)   & ---   & 1806 (289)     & ---  \\
& CO stretch       & 1798 & 1808 (10)     & 1809 (10)     & 1808 (9)    & ---   & 1806 (8)       & ---  \\
& CH sym. stretch  & 2739 & 2813 (74)     & 2814 (74)     & 2816 (76)   & ---  & 2592 (147)     & ---   \\
& CH asym. stretch & 2749 & 2813 (64)     & 2814 (65)     & 2816 (67)   & ---  & 2592 (157)     & ---  \\
& MAE              & ---  & 34            & 32            & 32          & ---         & 191            & ---         \\
\hline
HCOOH
& OCO bend       & 626  & 622 (3)    & 628 (2)       & 623 (3)  & 640 (14)     & 637 (12)       & 627 (1)      \\
& torsion      & 638  & 632 (6)       & 628 (11)      & 623 (16)    & 640 (1)      & 637 (1)        & 633 (6)      \\
& CH bend           & 1034 & 1043 (8)      & 1049 (15)     & 1046 (11)   & 1153 (118)   & 1168 (134)     & 1167 (133)   \\
& CO stretch          & 1111 & 1107 (5)      & 1068 (43)     & 1046 (66)   & 1153 (41)    & 1168 (56)      & 1167 (56)    \\
& OH bend      & 1253 & 1270 (17)     & 1256 (3)      & 1276 (23)   & 1206 (47)    & 1221 (32)      & 1286 (33)    \\
& CH bend          & 1394 & 1458 (64)     & 1518 (123)    & 1331 (63)   & 1386 (8)     & 1221 (173)     & 1405 (11)    \\
&   C=O stretch   & 1802 & 1820 (18)     & 1820 (17)     & 1820 (18)   & 1821 (19)    & 1820 (17)      & 1812 (10)    \\
& CH stretch       & 2906 & 2986 (79)     & 2986 (80)     & 2986 (79)   & 2779 (128)   & 2780 (126)     & 2530 (376)   \\
& OH stretch       & 3509 & 3500 (9)      & 3595 (87)     & 3604 (96)   & 3175 (334)   & 3076 (433)     & 2530 (978)   \\
& MAE              & ---  & 23            & 42            & 42          & 79           & 109            & 178          \\
\hline
\end{tabular}

\begin{tablenotes}
\footnotesize
\item[a] \nick{This analysis was not performed for the original TPB H$_2$CO trajectory because numerical instabilities occurred at longer times. The original TPB method was also numerically unstable for the H$_2$ and H$_2$O systems.}
\end{tablenotes}
\par\noindent{Alt text:} \textit{Table analyzing convergence of computed vibrational frequencies for H$_2$, H$_2$O, H$_2$CO, and HCOOH.}
\end{threeparttable}
\end{table*}

\newpage

\section{\MakeUppercase{Geometries}}

\noindent\textbf{HCN}

\begin{verbatim}
$molecule
0 1
N       0.0314554897     0.0000000000     0.0000000000
C       1.1872453288     0.0000000000     0.0000000000
H       2.1572991815     0.0000000000     0.0000000000
$end
\end{verbatim}

\noindent\textbf{HNC}

\begin{verbatim}
$molecule
0 1
C       1.1266466869     0.0000000000     0.0000000000
N      -0.0496579738     0.0000000000     0.0000000000
H      -0.9547020625     0.0000000000     0.0000000000
$end
\end{verbatim}

\noindent\textbf{FHF$^-$}

\begin{verbatim}
$molecule
-1 1
F       0.0000000000     0.0000000000    -1.1456813809
F       0.0000000000     0.0000000000     1.1456813809
H       0.0000000000     0.0000000000     0.0000000000
$end
\end{verbatim}

\noindent\textbf{H$_2$}

\begin{verbatim}
$molecule
0 1
H      -0.2812548842     0.0000000000     0.0000000000
H       0.2812548842     0.0000000000     0.0000000000
$end
\end{verbatim}

\noindent\textbf{ H$_2$O}

\begin{verbatim}
$molecule
0 1
O       0.6801139623     0.3972099168     2.7427695939
H       1.5427590058     0.3646621872     2.7170711076
H       0.4316670319    -0.2518821040     2.2301092985
$end
\end{verbatim}

\noindent\textbf{H$_2$CO}

\begin{verbatim}
$molecule
0 1
O       2.8199694258     0.5698762453     3.6232799137
C       1.6390008645     0.5169109265     3.3901151115
H       1.1919721501     1.0907627469     2.6524504061
H       0.9968175595    -0.1057999188     3.9127245686
$end
\end{verbatim}

\noindent\textbf{HCOOH}

\begin{verbatim}
$molecule
0 1
C      -0.6414393235     0.3130446546     2.2373247123
O       0.0392050442     0.1574030015     1.0840612904
O      -0.1351788690     0.4773256262     3.3164417302
H      -1.6390522598     0.2682512563     2.0400303027
H       0.8807854081     0.2013854614     1.2934319644
$end
\end{verbatim}

\noindent\textbf{oHBA}

\begin{verbatim}
$molecule
0 1
C -1.293251 1.270970 0.000000
C 0.032066 0.824049 0.000000
C 0.294115 -0.568710 0.000000
C -0.781038 -1.473582 0.000000
C -2.086523 -1.025436 0.000000
C -2.330627 0.354621 0.000000
O 1.019570 1.730158 0.000000
C 1.660600 -1.058238 0.000000
O 2.655089 -0.337865 0.000000
H -3.352319 0.715847 0.000000
H -0.564373 -2.536711 0.000000
H -2.911047 -1.726098 0.000000
H -1.480360 2.337035 0.000000
H 1.782954 -2.156428 0.000000
H 1.894815 1.229515 -0.000000
$end
\end{verbatim}

\noindent\textbf{BP(OH)$_2$}
\begin{verbatim}
$molecule
0 1
N       3.4755597082    -0.2805529509    -2.3013309785
C       4.4611179563    -0.4725697041    -4.6370620183
H       6.5025945376    -0.4994860990    -4.7747410280
C       2.9074879095    -0.6285910332    -6.7726775699
H       3.7447958181    -0.7816822548    -8.6357163266
C       0.2903263420    -0.5841317561    -6.4789402108
H      -0.9836391765    -0.6999790876    -8.0772960869
C      -0.7563742735    -0.3834015836    -4.0474463897
C       0.9233207992    -0.2324115573    -1.9517417032
C       0.0005479330    -0.0219111576     0.6452514743
N      -2.5516979386     0.0258264705     0.9949847120
O      -3.2905616700    -0.3376211558    -3.7214210750
C      -3.5371999729     0.2176102863     3.3307642243
C      -1.9835091800     0.3737515906     5.4663092531
C       1.6802966435     0.1292849829     2.7409056374
C       0.6336499597     0.3297173629     5.1724381437
H      -5.5786750191     0.2441999089     3.4685250059
H      -2.8207508980     0.5266322315     7.3293940116
H       1.9076620081     0.4457127972     6.7707454044
O       4.2145498905     0.0840500997     2.4149115522
H       4.5695479243    -0.0754452557     0.4238779449
H      -3.6449541116    -0.1776778141    -1.7302034035
$end
\end{verbatim}
\section{\MakeUppercase{Example input file for Q-Chem}}

\begin{verbatim}
$molecule
0 1
N       0.0314554897     0.0000000000     0.0000000000
C       1.1872453288     0.0000000000     0.0000000000
H       2.1572991815     0.0000000000     0.0000000000
$end
$rem
jobtype = opt
method = B3LYP
neo_epc               = epc172
NEO                   = true
SYM_IGNORE            = true
NO_REORIENT    TRUE
INPUT_BOHR            = false
BASIS                 = cc-pVDZ
THRESH 14
SCF_CONVERGENCE 10
max_scf_cycles 500
XC_GRID              	= 000099000590
NL_GRID              	= 000099000590
NEO_N_SCF_CONVERGENCE = 9
NEO_E_CONV            = 9
neo_basis             = PB4-D
NEO_VPP               = 0
NEO_SIMULTANEOUS_SCF  = true
NEO_PURECART = 1111
MEM_TOTAL 30000
GEOM_OPT_TOL_GRADIENT 1
GEOM_OPT_TOL_DISPLACEMENT 1
GEOM_OPT_TOL_ENERGY 1
$end
@@@
$molecule
    READ
$end
$rem
NEO                   = true
SYM_IGNORE            = true
NO_REORIENT    TRUE
INPUT_BOHR            = false
BASIS                 = cc-pVDZ
METHOD                = B3LYP
THRESH                = 14 
neo_epc               = epc172
SCF_CONVERGENCE       = 11
XC_GRID              	= 000099000590
NL_GRID              	= 000099000590
max_scf_cycles        = 300
NEO_N_SCF_CONVERGENCE = 9
NEO_E_CONV            = 9
neo_basis             = PB4-D
NEO_VPP               = 0
NEO_TDKS              = true
NEO_SIMULTANEOUS_SCF  = true
NEO_PURECART = 1111
$end
$neo_tdks
METHOD  = BO-Ehrenfest
DT      = 4
MAXITER = 2000
update_nuclei_every_n = 1
ctpb = true
tpb = true
deltaR = true
$end


\end{verbatim}
\renewcommand{\refname}{REFERENCES}
\bibliography{cite}